\documentclass[conference]{IEEEtran}
\IEEEoverridecommandlockouts

\usepackage{cite}
\usepackage{amsmath,amssymb,amsfonts}
\usepackage{graphicx}
\usepackage{textcomp}
\usepackage{xcolor}
\usepackage{float}
\usepackage{comment}
\def\BibTeX{{\rm B\kern-.05em{\sc i\kern-.025em b}\kern-.08em
    T\kern-.1667em\lower.7ex\hbox{E}\kern-.125emX}}
\usepackage{algorithm}

\usepackage{algpseudocode}
\usepackage{booktabs}
\usepackage{subcaption}
\usepackage{fancyhdr}
\usepackage{ulem}

\usepackage{amsthm}    
\newtheorem{definition}{Definition}
\newtheorem{example}{Example}

\newtheorem{theorem}{Theorem}
\newtheorem{lemma}{Lemma}   \newtheorem{proposition}{Proposition}

\DeclareMathOperator*{\argmax}{arg\,max}  

\renewcommand{\algorithmicrequire}{\textbf{Input:}}
\renewcommand{\algorithmicensure}{\textbf{Output:}}

\fancypagestyle{FirstPage}{

\lfoot{\footnotesize © 2022 IEEE.  Personal use of this material is permitted.  Permission from IEEE must be obtained for all other uses, in any current or future media, including reprinting/republishing this material for advertising or promotional purposes, creating new collective works, for resale or redistribution to servers or lists, or reuse of any copyrighted component of this work in other works.}

}

\begin{document}

\title{Statistical Quantification of Differential Privacy: \\A Local Approach}

\author{\IEEEauthorblockN{Önder Askin}
\IEEEauthorblockA{\textit{}
\textit{Ruhr-University Bochum}\\
oender.askin@rub.de}
\and
\IEEEauthorblockN{Tim Kutta}
\IEEEauthorblockA{\textit{}
\textit{Ruhr-University Bochum}\\
tim.kutta@rub.de}
\and
\IEEEauthorblockN{Holger Dette}
\IEEEauthorblockA{\textit{}
\textit{Ruhr-University Bochum}\\
holger.dette@rub.de}
}

\maketitle

\begin{abstract}
In this work, we introduce a new approach for statistical quantification of differential privacy in a black box setting. We present estimators and confidence intervals for the optimal privacy parameter of a randomized algorithm $A$, as well as other key variables (such as the “data-centric privacy level”). Our estimators are based on a local characterization of privacy and in contrast to the related literature avoid the process of “event selection” - a major obstacle to privacy validation. 
This makes our methods easy to implement and user-friendly. We show fast convergence rates of the estimators and asymptotic validity of the confidence intervals. An experimental study of various algorithms confirms the efficacy of our approach.
\end{abstract}

\begin{IEEEkeywords}
Differential privacy, data-centric privacy, local estimators, confidence intervals
\end{IEEEkeywords}

\section{Introduction}

\thispagestyle{FirstPage}

Since its introduction in the seminal work of \cite{Dwork2006}, the concept of \textit{Differential Privacy} (DP) has become a standard tool to assess information leakage in data disseminating procedures. DP characterizes how strongly the output of a randomized algorithm is influenced by any one of its inputs, thus quantifying the difficulty of inferring arguments (i.e., user information) from algorithmic releases. 

To formalize this situation, we consider a database $x = (x(1), \cdots, x(m))$ where each data point $x(i)$ takes values in a set $ \mathcal{D}$ and corresponds to the data provided by the $i$th individual among $m$ users. 
Furthermore, we introduce the notion of \textit{neighboring} or \textit{adjacent} databases, that is databases that only differ in one component. Mathematically, we can express neighborhood of $x,x'$ by unit Hamming distance $d_H(x,x')=1$, where the Hamming distance is defined as follows:
\begin{align*}
d_H(x,x') := \vert \{ 1 \leq i \leq m: x(i) \neq x'(i) \} \vert.
\end{align*}

\begin{definition} \label{def_privacy}
An Algorithm $A$ is called $\epsilon$-differentially private for some $\epsilon>0$, if for any two neighboring databases $x,x'$ and any measurable event $ E $ the inequality
\begin{align}
\mathbb{P}(A(x) \in E) \leq e^{\epsilon} \, \mathbb{P}(A(x') \in E) \label{Def_DP}
\end{align}
holds.
\end{definition}
Definition \ref{def_privacy} demands that \eqref{Def_DP} holds for all measurable events $E$, but what constitutes a measurable event depends on the output space $\mathcal{Y}$ of the randomized algorithm $A$. If $\mathcal{Y}$ is discrete (in particular if $|\mathcal{Y}|<\infty$) we require that \eqref{Def_DP} holds for all events in the power set $\mathcal{P}(\mathcal{Y})$. If however $A$ has outputs in a continuum (e.g., $\mathcal{Y} = \mathbb{R}
^d$), then \eqref{Def_DP} has to hold for all Borel sets. In both cases, the collection of all measurable events is large and complex, which is an important obstacle in the practical validation of DP as we will discuss below. 

The privacy parameter $ \epsilon$ in Definition \ref{def_privacy} quantifies the information leakage of $A$, where small values correspond to small leakage (and thus high privacy). Hence, deploying differentially private algorithms with appropriate $\epsilon$ provides users with strong privacy guarantees regarding their data. 
Aware of these properties, there has been an increased interest in and deployment of differentially private algorithms by companies that handle large amounts of data (such as Google \cite{Erlingsson2014}, Microsoft \cite{Bolin2017} and Uber \cite{Johnson2018}), as well as government agencies such as the US Census Bureau \cite{Abowd2018}.
However, in practice it is often unclear whether an algorithm satisfies DP and if so, for which parameter $\epsilon$. It is therefore important and the main objective of this work to develop procedures by which we can ascertain the level of privacy afforded by a given algorithm. We will focus on “pure” DP as defined in \eqref{Def_DP} in this work and refer readers interested in “approximate differential privacy” to \cite{Barthe2013,Barthe2014,Barthe2016,Liu2019,Barthe2020}. 

\subsubsection*{\textbf{Related work}}
A number of languages and verification tools have been devised to validate differential privacy where possible and discard it where not (see among others \cite{Reed2010, Gaobardi2013,Barthe2016b,Hsu2017, Zhang2017,Kifer2019,Pierce2020,CheckDP}). Many of these approaches are designed specifically for developers and require knowledge of the inner structure of the algorithm in question. In contrast, in this paper, we want to investigate a black box scenario where we have little to no knowledge of the algorithm's design and have to rely solely on output samples. This scenario can occur naturally when third parties are entrusted with validating the privacy claims of a data collector. In this situation, skeptical users and agencies can confirm the privacy of a given algorithm, while the data collector does not have to reveal his (proprietary) source code and algorithm design. However, black box methods can also be valuable in settings where an algorithm is known but so complex, that focusing on its outputs is preferable. In any case, a procedure tailored to this scenario covers a wide range of algorithms with few requirements, which is a desirable feature in a validation scheme.

Relying solely on algorithmic outputs warrants a statistical approach and such methods are pursued in \cite{StatDP}, built directly on Definition \ref{def_privacy}. For a fixed triplet $(x,x',E)$ consisting of neighboring databases $x,x'$ and an event $E$, the privacy condition in \eqref{Def_DP} can be construed as a statistical hypothesis that needs to be checked. Given a preconceived privacy parameter $\epsilon_0 > 0$, candidate triplets
are generated and a binomial statistical test is employed to find a counterexample $(x_0,x'_0,E_0)$ that violates the privacy condition \eqref{Def_DP}.
These counterexamples expose faulty, non-private algorithms in a fast and practical manner and hint at potential weaknesses in the algorithm's design.

A related, but distinct approach is the examination of lower bounds for differential privacy \cite{DP-Finder}. Here, privacy violations are determined with the help of the “privacy loss”, which is defined for any triplet $(x,x',E)$ as
\begin{align}
L_{x,x'}(E) := \Big| \ln \big( \mathbb{P}(A(x) \in E ) \big) - \ln \big( \mathbb{P}(A(x') \in E ) \big) \Big| \label{loss_events}.
\end{align}
We interpret $\infty-\infty := 0$ to account for events with $0$ probability.
In line with Definition \ref{def_privacy}, an algorithm $A$ satisfies $\epsilon$-DP if and only if $L_{x,x'}(E)\le \epsilon$ for all permissible triplets. Thus, computing privacy violations $L_{x,x'}(E)$ for different triplets naturally provides lower bounds for $\epsilon$. Note that in this context, privacy violations and loss are used constructively to gather information about the privacy parameter. We also want to point out that this approach can be adapted to counterexample generation, if for some predetermined $\epsilon_0$ a triplet  $(x_0,x'_0,E_0)$ is found s.t. $L_{x_0,x_0'}(E_0)>\epsilon_0$. However, lower bounds are somewhat more flexible, because they do not require some hypothesized $\epsilon_0$ in the first place.  

Even though \cite{StatDP} and \cite{DP-Finder} provide effective tools for privacy validation, they are not entirely compatible with our black box assumption. While the binomial test in \cite{StatDP} by itself requires little knowledge of $A$, the larger scheme, within which it is embedded, is designed to also consider the algorithm's program code. A symbolic execution of that code can be performed to  facilitate the detection of counterexamples. Therefore, this approach is also labeled  \textit{semi-black-box} by its authors \cite{StatDP}. Even less compatible with the black box regime, the approach in \cite{DP-Finder} requires access to the program code of algorithm $A$ in order to alter it in ways that produce a differentiable surrogate function for $L_{x,x'}$. Numerical optimizers can then be deployed to find triplets that yield high privacy violations.

A more recent method to quantify DP is the DP-Sniper algorithm, developed in \cite{DP-Sniper}. For fixed databases $x$ and $x'$, DP-Sniper creates an event $E^*$ which approximately maximizes \eqref{loss_events} and then derives a statistical lower bound for $L_{x,x'}(E^*)$. To construct $E^*$, a machine learning classifier is employed that approximates the posterior probability of $x$ given an output of $A$. Intuitively, $E^*$ then consists of all those outputs, that are expected to be  generated by $A(x)$ rather than $A(x')$ with high certainty. 
The classifiers used are logistic regression (a one-layer neural network) and a small neural network (two hidden layers). 
Both choices yield relatively simple parametric models for the posterior, where the classifier based on logistic regression corresponds to a  linear decision rule.
The successful maximization of $L_{x,x'} $ in \cite{DP-Sniper} then presupposes that the true (and unknown) posterior distribution belongs to 
one of these classes.
Naturally, such a parametric assumption limits the scope of  theoretical performance guarantees and is difficult to reconcile with a black box setting, where a non-parametric statistical procedure would be more fitting.

\subsubsection*{\textbf{The problem of event selection}}

As we have seen above, statistical validation of DP rests on finding a triplet $(x,x',E)$ that provokes a high privacy violation. This task is typically split into two separate parts: First, finding databases $x,x'$ such that the loss $L_{x,x'}(E)$ is large for some event $E$ and, second, finding this very event. Even though both problems are non-trivial, the greater challenge lies in the latter one, the \textit{event selection} (see \cite{DP-Sniper}).

Starting with the space of potential events, we observe that if $\mathcal{Y}$ consists of a finite number of output values, the number of measurable events grows exponentially in $|\mathcal{Y}|$ with $|\mathcal{P}(\mathcal{Y})| = 2^{\vert \mathcal{Y} \vert}$. This makes evaluating $L_{x,x'}$ on all potential events $E$ impractical even if $\vert \mathcal{Y} \vert < \infty $, and the task becomes impossible if $\mathcal{Y}$ is a continuum. Therefore, a prior restriction is necessary to narrow down candidate events. In related works, this process is guided by heuristics \cite{StatDP} or parametric assumptions \cite{DP-Sniper}. However, such approaches are in tension with a genuine black-box scenario, as they do not offer a template that generalizes to any given algorithm.

Event selection also poses a challenge from a learning perspective. Approximating the objective function $L_{x,x'}$ over a class of events entails a bias-variance trade-off: Here a larger class of events may help to find higher privacy violations, but it also requires higher sampling efforts to ensure uniform approximation. 
Furthermore, it can be difficult to control the optimization error, as the objective function $L_{x,x'}$ eludes classical numerical treatment (it does not satisfy continuity, differentiability, etc.).

As a consequence of these difficulties, we propose an alternative route to assess  DP in this work. Rather than searching for vulnerable events, we approximate the maximum $\sup_E L_{x,x'}(E)$ directly using a \textit{local loss function}  (see Section III). By circumventing event selection, we can effectively reduce complexity and algorithmic effort to quantify the privacy level of a given algorithm (see Section 4 and 5).

\subsubsection*{ \textbf{Data-specific privacy violations}}

In this work, a central object of interest is the quantity
\begin{align} 
\epsilon_{x,x'} & := \sup_E L_{x,x'}(E)
\label{lower_bound}
\end{align}
which we call \textit{data-specific privacy violation} in $x$ and $x'$. Recalling \eqref{loss_events}, we observe that $\epsilon_{x,x'}$ indicates to which extent  the algorithm outputs are indistinguishable  for a fixed pair of databases $x$ and $x'$. Note that $A$ satisfies $\epsilon_0$-DP if and only if $\epsilon_{x,x'} \le \epsilon_0$ for all pairs of adjacent databases $(x,x')$. Thus, we define the smallest parameter $\epsilon$, for which $\epsilon$-DP still holds as
\begin{align} \label{Eq_max_violation}
\epsilon := \sup\limits_{x,x': \; d_H(x,x') = 1} \epsilon_{x,x'},
\end{align}
and note that $\epsilon$ is optimal in the sense that privacy guarantees below $\epsilon$ are not feasible, while any $\epsilon_0 > \epsilon$ underestimates the privacy level that is actually achievable.

We refer to $\epsilon$ as the \textit{global privacy parameter} which, in light of  identity  \eqref{Eq_max_violation}, only provides a “worst-case” guarantee for privacy leakage of any pair $x,x'$. In contrast, the precise amount of privacy leakage associated with $x$ and $x'$ is captured by $\epsilon_{x,x'}$, which is potentially much smaller than $\epsilon$. The data-specific privacy violations comprise more granular  information that we utilize to examine the following privacy aspects:

First, each $\epsilon_{x,x'}$ constitutes a lower bound of $\epsilon$. Because $L_{x,x'}(E) \leq \epsilon_{x,x'}$ holds for all events $E$, these lower bounds are at least equally and potentially even more powerful than the ones derived in prior work. Lower bounds in themselves are useful, as they can help expose faulty algorithms \cite{DP-Finder} and narrow down the extent to which a given algorithm can be private at all \cite{DP-Sniper}. This ultimately provides us with a better understanding of the global privacy parameter $\epsilon$.

Secondly, data-specific privacy violations can be used to infer the \textit{data-centric privacy level} for select databases. More precisely, suppose that a curator has gathered a database $x$ and is interested in the amount of privacy conceded specifically to the individuals with data in $x$. The maximum privacy violation associated with $x$ is obtained by forming the supremum over all data-specific privacy violations in its neighborhood, that is
\begin{align}
    \label{Eq_def_epsx}
\epsilon_x := \sup\limits_{x': \; d_H(x,x') = 1} \epsilon_{x,x'}.
\end{align}
Graphically speaking, $\epsilon_x$ is the maximum privacy loss attained on a unit sphere around $x$  (with regard to $d_H$). It also constitutes the maximum privacy loss any individual represented in $x$ has to at most tolerate (thus, it has also been studied in the context of "individual DP" \cite{IndividualDP}). Evidently, we have $\epsilon_x \leq \epsilon$ for all databases $x$ and we will see later on that the data-centric privacy level $\epsilon_x$ can be considerably smaller than the global privacy guarantee $\epsilon$ (see Section 5).

The relation between specific databases and privacy has been previously studied in the context of sensitivity \cite{Nissim2007}. Given a function $F$ that operates on databases $x$, one can achieve DP by adding noise proportional to the global sensitivity $\triangle_F$ of $F$ to its output $F(x)$. \cite{Nissim2007} observe that the local sensitivity $\triangle_F(x)$ of $F$ around a fixed database $x$ can be considerably smaller than $\triangle_F$, allowing for, in principle, less noise and higher accuracy. The local sensitivity of $F$ is then leveraged to arrive at the notion of “smooth sensitivity”, which admits lower levels of noise than $\triangle_F$ and can be analytically determined for some statistically relevant functions.

In the presence of only black box access to the target function $F$, \cite{Nissim2007} avoid computing the sensitivity of $F$ directly and instead resort to assessing the sensitivity of an aggregation function operating on outputs of $F$. In contrast, \cite{Rubinstein2017} propose an approach that provides direct sensitivity estimates of the target function $F$ that can be used in the privatization process. As a sampling-based black box method, the approach put forward in \cite{Rubinstein2017} shares some similarities with our methodology, but also comes with marked differences. The methods in \cite{Rubinstein2017} assist directly in the design of algorithms that conform to a relaxed version of DP, namely random differential privacy \cite{Hall2013}. We, on the other hand, develop statistical methods that assess “pure” DP and, given a randomized algorithm, determine the privacy level $\epsilon_x$ attached to a  database $x$ in retrospect.

\subsubsection*{\textbf{This work}} Statistically, our approach is based on novel estimators $\hat \epsilon_{x,x'}$ for the data-specific privacy violation $\epsilon_{x,x'}$. In view of the identities \eqref{Eq_max_violation} and \eqref{Eq_def_epsx}, such estimates are natural building blocks for the assessment of the global privacy parameter $\epsilon$ or its data-centric version $\epsilon_x$. Contrary to the related literature, our estimators do not maximize an empirical version of the loss $L_{x,x'}$, but approximate the supremum $\epsilon_{x,x'}$ directly, thus avoiding the pitfalls of event selection (see previous part). Mathematically, these estimates rest on a “local” version of the privacy loss discussed in Section \ref{Sec_3}. Besides estimators, we present new tools of statistical inference: In Section \ref{Sec_4} we devise the MPL (Maximum Privacy Loss) algorithm, which generates one-sided confidence intervals $[LB, \infty)$ for the privacy parameters $\epsilon$ and $\epsilon_x$ respectively. In this situation, $LB$ is a statistical lower bound (i.e., it holds with a high degree of certainty) and approximates the true parameter with increasing sample size. In particular, if MPL is applied to the quantification of $\epsilon$ and outputs $LB$, the user can be confident that algorithm $A$ is at best $LB$-differentially private. In Section \ref{Sec_5} we confirm these findings via experiments.

\subsubsection*{\textbf{Main contributions}}

We give a brief summary of our main contributions:

\begin{itemize}
    \item A fully statistical black box procedure for the quantification of DP (without parametric assumptions).
    \item A flexible approach based on data-specific privacy violations $\epsilon_{x,x'}$ as building blocks.
    \item New estimators $\hat \epsilon_{x,x'}$ for the data-specific privacy violation that circumvent the problem of event selection and are proved to converge at a fast rate. 
    \item The MPL algorithm that outputs a confidence interval for $\epsilon$ (or $\epsilon_x$), which demonstrably includes the parameter of interest with approximate level of confidence.
    \item A practical evaluation and validation of our methods. 
\end{itemize}

\section{Statistical preliminaries}

In this section, we review the statistical concepts of \textit{confidence intervals} and \textit{kernel density estimation}, which serve as technical background for the remainder of this paper. Readers who are only interested in discrete algorithms can omit Section \ref{Subsec_2_B}. 

\subsection{Confidence Intervals} \label{Subsec_2_A}

A confidence interval is a statistical method to localize a parameter of a probability distribution with a prescribed level of certainty. More concretely, consider a sample of $n$  observations $X_1,..,X_n$ (random variables),  following an unknown distribution $P$. If a user is interested in a parameter $\theta  = \theta(P)$ derived from $P$ (e.g. the expectation $\theta:=\mathbb{E}_P X_1$), the sample of observations can be used to approximately locate $\theta$ in an interval $\hat I(X_1,...,X_n) \subset \mathbb{R}$. Notice that the term \textit{confidence interval} usually refers to both the output $\hat I(X_1,...,X_n)$, which is an interval  determined by the data, and the underlying algorithm $\hat I (\cdot)$ itself. Given the randomness in the data, there is always a risk of mislocating $\theta$, i.e. that $\theta \not \in \hat I(X_1,...,X_n)$. However, confidence intervals are constructed to guarantee $\theta \in \hat I(X_1,...,X_n)$ with a prescribed probability (level of confidence). To be more precise, $\hat I(\cdot )$ has an additional input parameter $\alpha \in (0,1)$, such that the \textit{confidence level} $1-\alpha$ holds:
\begin{equation} \label{Eq_def_conf_int}
\mathbb{P}(\theta   \in \hat I_\alpha(X_1,...,X_n)) =1- \alpha,
\end{equation}
 where typically $\alpha \in \{0.1, 0.05, 0.01\}$. Notice that the choice of $\alpha$ entails a trade-off: On the one hand a smaller $\alpha$ provides the user with higher certainty that actually $\theta  \in \hat I_\alpha(X_1,...,X_n)$, but on the other hand it translates into a wider confidence interval, which means less precision with regard to the location of $\theta$. Besides the choice of $\alpha$, the sample size $n$ affects the width of the confidence interval, with larger $n$ leading to narrower intervals.\\
In order to construct a confidence interval $\hat I_\alpha$ s.t. \eqref{Eq_def_conf_int} holds, it is necessary to have prior knowledge about the underlying distribution of the data sample $X_1,..., X_n$. For instance, it may be known that the sample comes from a normal distribution, with unknown mean and variance, and we want to give a confidence interval for the mean. In this situation, parametric statistical theory equips the user with standard tools to construct $\hat I_\alpha$ (see \cite{bickel2015}).\\ Yet in many cases such prior knowledge about the data is not feasible and therefore a weaker requirement than \eqref{Eq_def_conf_int} is formulated: It states that the confidence level $1-\alpha$ is approximated with increasing precision, as $n$ grows larger, or mathematically speaking 
\begin{equation} \label{Eq_def_as_conf_int}
\lim_{n \to \infty}\mathbb{P}(\theta   \in \hat I_\alpha(X_1,...,X_n)) = 1-\alpha.
\end{equation}
If \eqref{Eq_def_as_conf_int} is satisfied, we call $\hat I_\alpha$ an \textit{asymptotic confidence interval with confidence level $1-\alpha$}. The advantages of asymptotic confidence intervals are their flexibility and robustness against deviations from a presumed distribution. Common approaches to prove asymptotic confidence levels include asymptotically normal estimators, as well as the delta method for differentiable statistics. For details on asymptotic statistical theory, we refer the interested reader to the monograph of \cite{vandervaart1996}.

\subsection{Kernel density estimation} \label{Subsec_2_B}
Kernel density estimation is a method to estimate the unknown distribution of a data sample $X_1,...,X_n$ on $\mathbb{R}^d$. 
It can be thought of as the creation of a smoothed, normalized histogram, where the jumps between the bins are interpolated continuously (for an introduction see \cite{scott}). This procedure is often preferred to a traditional histogram, particularly if the data sample is distributed according to a continuous density $f$ on $\mathbb{R}^d$ (we write $X_1,...,X_n \sim f$).  \\
More precisely, let $K: \mathbb{R}^d \to \mathbb{R}$ be a continuous, non-negative function, such that $\int_{\mathbb{R}^d} K(u) du =1$. We call $K$ a kernel and define the \textit{kernel density estimator} (KDE) $\tilde f$ for $f$ pointwise as 
\begin{equation} \label{Eq_def_KDE}
    \tilde f(t) := {1 \over nh^d} \sum_{i=1}^n K \bigg( {t-X_i \over h} \bigg), \quad t \in \mathbb{R}^d,
\end{equation}
where $h>0$ is the \textit{bandwidth}, analogue to the bin-width in a histogram. For details on kernel density estimators as well as generalizations such as multidimensional bandwidths, we refer to \cite{Gramacki}. 
As the number of observations $n$ increases, the convergence speed of $\tilde f$ to $f$ depends on three distinct factors: First the smoothness of the true density $f$, secondly
an adequate choice of the kernel $K$ and thirdly the bandwidth $h$.\\
To quantify smoothness we require $f$ to be \textit{Hölder continuous}, i.e. for some $\beta \in (0,1]$  and $C>0$ it holds that 
\begin{equation}\label{Eq_Hoelder}
    |f(t)-f(s)| \le C |t-s|^\beta, \quad \forall t,s \in \mathbb{R}^d~,
\end{equation}
where $|\cdot|$ denotes the Euclidean norm. Notice that $\beta=1$ corresponds to the well known \textit{Lipschitz continuity}, which is satisfied by the densities corresponding to the Laplace, Gaussian and versions of the Exponential Mechanism. We also point out that a density which satisfies Hölder continuity for one $\beta>0$ is  Hölder continuous for any other $\beta' \in (0, \beta]$. \\
The choice of the kernel $K$ is a relatively simple task: To attain optimal convergence speed, $K$ has to fulfill certain regularity properties (K1) and (K2), that we make precise in Appendix B. From now on we will always assume that $K$ conforms to these assumptions. We point out that both of them are satisfied by all commonly used  kernels (in particular by the Gaussian kernel, that we use in our experiments).

\begin{figure}
\centering
\includegraphics[width=0.8\linewidth,height=140pt]{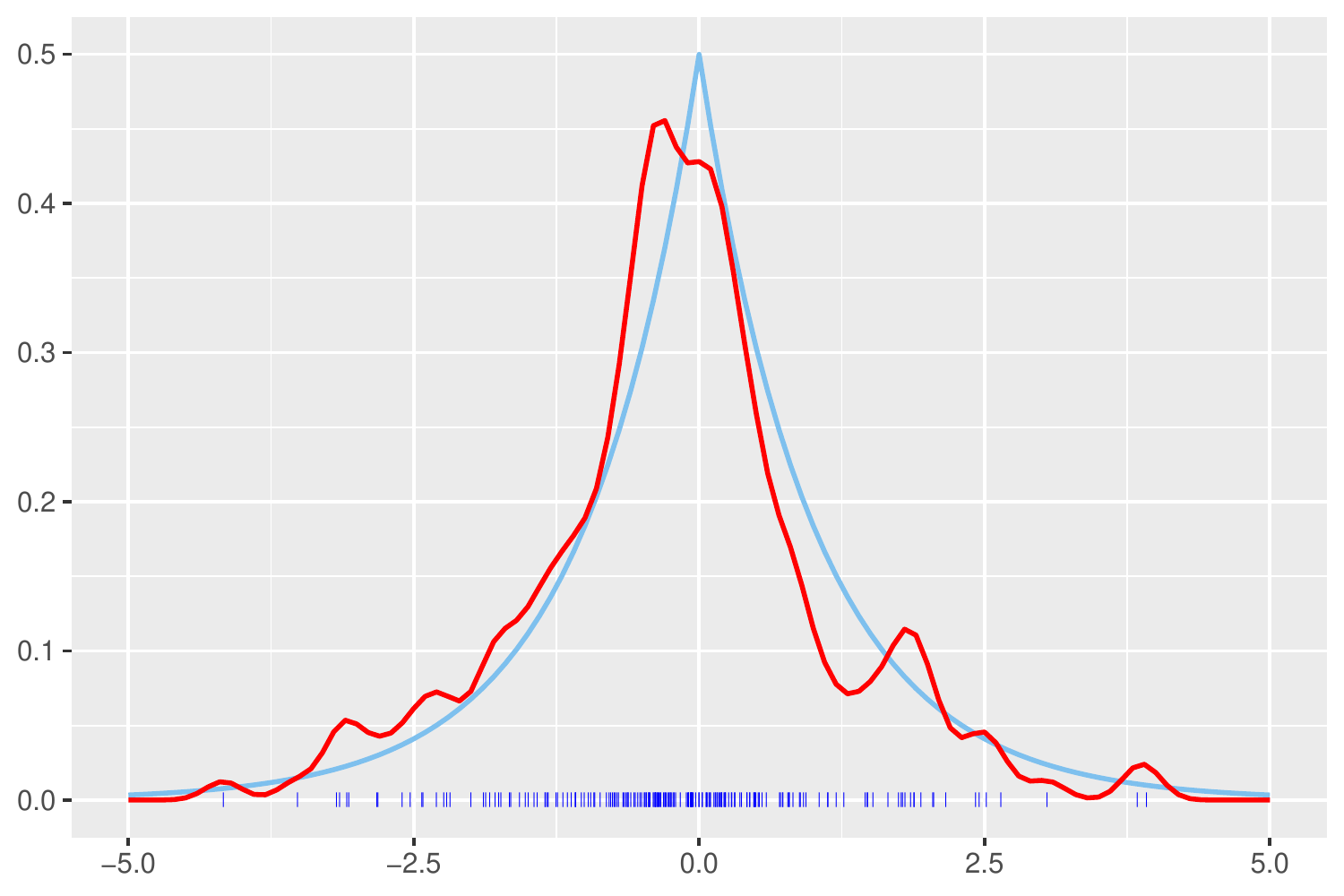}
\caption{ Centered Laplace density (light blue) and kernel density estimate (red) for $N=200$, with Gaussian kernel. On the $x$-axis we have plotted the observations $X_1,...,X_{200}$ (dark blue).\\[-5ex]}
\end{figure}

Finally, the choice of the bandwidth $h$ should depend on the smoothness level $\beta$ of $f$, as well as the sample size $n$. More precisely, it can be shown that 
\begin{equation} \label{Eq_KDE_rate_1}
    \sup_{t \in \mathbb{R}^d} |\tilde f(t)-f(t)| = \mathcal{O}_P \Big( h^\beta + \sqrt{\frac{\ln(n)}{h^d n}}\Big),
\end{equation}
which implies for the specific choice
 $h= \mathcal{O}(n^{-\frac{1}{2\beta+d}})$ 
\begin{equation} \label{Eq_KDE_rate_2}
    \sup_{t \in \mathbb{R}^d} |\tilde f(t)-f(t)| = \mathcal{O}_P \Big( \sqrt{\ln(n)} n^{-\frac{\beta}{2\beta+d}}\Big).
\end{equation}
 Notice that this $h$ minimizes the error rate (except for log-terms).
For details on convergence rates in density estimation see  \cite{jiang2017} and for a definition of the stochastic Landau symbol $\mathcal{O}_P$ we refer to the Appendix A.\\
 In practical applications the true smoothness $\beta$ and hence the optimal bandwidth is unknown and therefore data-driven procedures, such as cross validation, are used to determine it. For details on bandwidth selection, see \cite{Gramacki}. \\
In the subsequent discussion, we consider log-transformed density estimators. These objects are potentially unstable for arguments where the true density $f$ is close to $0$, because small errors in the estimate of $f$ translate into great errors in the logarithm. For this reason, we define the truncated KDE pointwise in $t$ as 
$$
\hat f (t) := \tilde f(t) \lor \tau,
$$
where “$a \lor b$” denotes the maximum of two numbers $a, b \in \mathbb{R}$ and $\tau>0$ is a user-determined floor. In Section \ref{Sec_4} we discuss how to choose $\tau$ dependent on $n$ and $\beta$. The construction of the truncated KDE is described in Algorithm \ref{alg}. 

\begin{algorithm}
	\caption{Truncated kernel density estimator} \label{alg}
	\small
	\algorithmicrequire \; \parbox[t]{\dimexpr\linewidth-\algorithmicindent}{
           data sample $X=(X_1,...,X_n)$, evaluation point $t$, \\ bandwidth $h$, kernel function $K$, floor $\tau$
        } \\
	\begin{algorithmic}[1]
	\Function{TKDE}{$ X, t, h, K, \tau$}\\
	    $out = 0$
		\For {$i=1,2,\ldots,n$}
				\State $out = out+ K((t-X_i)/h)$
		\EndFor \\
	$out = out/(n h^d)$\\
		\Return $out \lor \tau$
	\EndFunction
	\end{algorithmic} 
\end{algorithm}

\section{Differential Privacy as a local property} \label{Sec_3}

As we have seen in our Introduction, $\epsilon$-DP means that for any neighboring databases $x,x'$ the bound
\begin{equation} \label{Eq_data_specific}
\epsilon_{x,x'}=
\sup_E L_{x,x'}(E) 
\le \epsilon
\end{equation}
holds, where 
the loss $L_{x,x'}$ 
is 
defined in \eqref{loss_events}. 
Thus, in principle, validating DP requires the calculation of $L_{x,x'}(E)$ for any measurable event $E$, a problem that is intractable from a practical perspective  given the complexity of the space of measurable events (see Introduction). 
We can, however, drastically reduce the effort of \textit{event selection} in the supremum by exploiting that differential privacy is an inherently \textit{local property}, i.e. that the level of privacy is determined by the loss on small events. 
To get an intuition of this point, consider an event $E$ that can be decomposed into the disjoint subsets $E_1$ and $E_2.$ It is a simple exercise to show that
$$
L_{x,x'}(E) \le \max \{L_{x,x'}(E_1), L_{x,x'}(E_2)\}.
$$
In this sense going from larger to smaller events increases the privacy loss and thus gets us closer to $\epsilon_{x,x'}$. Iterating this process suggests that we should look at “the smallest events possible”, which are single points. So we expect that ultimately
\begin{align} \label{Eq_intuition}
\epsilon_{x,x'} \approx \sup_{t \in \mathcal{Y}}|L_{x,x'}(\{t\})|.
\end{align}
Admittedly, this statement is not formally correct for all algorithms, but we will make it rigorous for certain classes of algorithms in the course of this section. Compared with the supremum over all measurable events in \eqref{Eq_data_specific}, the expression in \eqref{Eq_intuition} is more convenient, because single points are easy to handle. We will explore this advantage in detail at the end of this section. 

We now begin our formal discussion by specifying two classes of algorithms that are considered throughout this work: discrete and continuous ones. \\
We call an algorithm $A$ that maps a database $x$ to random values in either a finite or a countably infinite set $ \mathcal{Y}$ a \textit{discrete algorithm}. Without loss of generality, we will assume that $\mathcal{Y} \subset \mathbb{N}$. Moreover, we call the corresponding  probability function $f_x: \mathcal{Y} \to [0,1]$ defined as
\begin{align}
f_x(t) := \mathbb{P}(A(x) =t), \quad \forall t \in \mathcal{Y}  \label{Def_discrete_density}
\end{align}
the \textit{discrete density} of $A$ in $x$. With this notation we can write for any $E \subset \mathcal{Y}$
\begin{equation} \label{Eq_def_discr_dens}
\mathbb{P}(A(x) \in E) = \sum_{t \in E}f_x(t).
\end{equation}
Examples of discrete algorithms include Randomized Response \cite{Kairouz2016}, Report Noisy Max \cite{Dwork2014} and the Sparse Vector Technique \cite{Lyu2017}.\\
Next, suppose that $\mathcal{Y} = \mathbb{R}^d$. We say that $A$ is a \textit{continuous algorithm}, if for any database $x$, $A(x)$ has a continuous density $f_x: \mathbb{R}^{d} \to \mathbb{R}$, such that for any Borel measurable event $E$ 
$$
\mathbb{P}(A(x) \in E) = \int_E f_x(t) dt.
$$
Typical examples of continuous algorithms are, as mentioned before, the Laplace \cite{Dwork2014}, the Gaussian \cite{Dwork2014} and versions of the Exponential Mechanism \cite{Talwar2007}.
We want to highlight that in this definition the requirement of continuous densities on the whole space $\mathbb{R}^d$ is only made for convenience of presentation and can be relaxed to densities on subsets, e.g., $[0, \infty) \subset \mathbb{R}$ in the case $d=1$. 
Notice that for continuous algorithms \eqref{Eq_intuition} is technically invalid because $L_{x,x'}(\{t\})=0$ for any point $t$.  However, it is possible to preserve the idea of \eqref{Eq_intuition} by reformulating it in terms of continuous densities (see Theorem \ref{theorem_local_property}).

Given the above definitions, the distribution of an algorithm $A$ can be thoroughly characterized by its densities and we use the notation  $A(x) \sim f_x$
throughout this paper.  In the following theorem, we give a mathematically rigorous  version of \eqref{Eq_intuition}. Variants of this theorem can be encountered in the DP literature and the inequality “$\le$” in \eqref{Loss_local} is frequently used in privacy proofs. However, the exact identity in \eqref{Loss_local}
is not trivial and therefore worked out here explicitly.

\begin{theorem} \label{theorem_local_property}
Given a discrete or continuous algorithm $A$ with $A(x) \sim f_x$ and $A(x') \sim f_{x'}$ we have 
\begin{align} 
\epsilon_{x,x'} = \sup_{t \in \mathcal{Y}} \big| \ln (f_x(t)) - \ln (f_{x'}(t)) \big|, \label{Loss_local}
\end{align}
where $\infty-\infty:= 0$.
\end{theorem}

\begin{IEEEproof} 
We first consider the discrete setting: In order to show “$\ge$” we notice that for all $t \in \mathcal{Y}$ 
$$
L_{x,x'}(\{t\})= \big| \ln (f_x(t)) - \ln (f_{x'}(t)) \big|.
$$
Recall that
$
\epsilon_{x,x'} = \sup_{E} | L_{x,x'}(E)  |.
$
Here the supremum is taken over all elements $E$ of the power set $\mathcal{P}(\mathcal{Y})$ (which includes in particular sets with only one element) and this directly implies “$\ge$”.\\
The proof of “$\le$” follows by standard techniques. We fix a set $E \subset \mathcal{Y}$
and rewrite $L_{x,x'}(E)$ using \eqref{Eq_def_discr_dens}, s.t.
\begin{align}
    L_{x,x'}(E) = \Big|\ln \Big( \frac{\sum_{t \in E} f_x(t)}{\sum_{t \in E} f_{x'}(t) } \Big)\Big|. \label{sum_quotient}
\end{align}
Without loss of generality, we assume that the numerator is greater than the denominator and we can therefore drop the absolute value. Now the inner fraction can be upper bounded as follows:
\begin{align*}
    \frac{\sum_{t \in E} f_x(t)}{\sum_{t \in E} f_{x'}(t) } \le \frac{\sum_{t \in E} f_{x'}(t)[f_x(t)/f_{x'}(t)]}{\sum_{t \in E} f_{x'}(t) } \le \sup_{t \in \mathcal{Y}} \frac{f_{x}(t)}{f_{x'}(t)}.
\end{align*}
Taking the logarithm on both sides and the supremum over all $E$ on the left maintains the inequality, showing “$\le$”. \\
Moving to continuous algorithms, we notice that the proof of “$\le$” follows along the same lines as for the discrete case and is therefore omitted (one simply has to replace all the sums by integrals). \\
To prove “$\ge$” we first observe that a probability density in $t$ gives the probability of a very small region around $t$. More precisely it can be expressed as follows
\begin{align*} 
    f_x(t) = \lim_{\delta \to 0} \frac{\mathbb{P}(A(x) \in U_\delta(t))}{vol(U_\delta(t))},
\end{align*}
where $U_\delta(t):=\{s \in \mathcal{Y}: |t-s|\le \delta \}$ and  $vol()$ denotes the $d$-dimensional volume. The identity is a special case of Theorem 6.20 (c) in \cite{Knapp2005}. The same statement holds for $x'$ instead of $x$ and we can use that to get
\begin{align*}
    \frac{f_x(t)}{f_{x'}(t)} = \lim\limits_{\delta \to 0} \frac{\mathbb{P}(A(x) \in U_\delta(t))}{\mathbb{P}(A(x') \in U_\delta(t))} \leq \sup\limits_{E} \frac{\mathbb{P}(A(x) \in E)}{\mathbb{P}(A(x') \in E)}
\end{align*}
for any $t \in \mathcal{Y}$. Taking the logarithm on both sides and the supremum over $t$ on the left preserves the inequality. Recalling \eqref{lower_bound}, this  implies $  \sup_{t \in \mathcal{Y}} \big| \ln (f_x(t)) - \ln (f_{x'}(t)) \big|\le \epsilon_{x,x'}$, which proves the theorem. 
\end{IEEEproof} 
$ $

Theorem \ref{theorem_local_property} allows us to characterize DP of an algorithm $A$ by the absolute log-difference of the algorithm's densities. For ease of reference we define this difference, the \textit{loss function}, explicitly as
\begin{align} \label{Eq_def_loss_function}
 \ell_{x,x'}(t) := \big| \ln (f_x(t)) - \ln (f_{x'}(t)) \big|. 
\end{align}
This definition admits the restatement of Theorem \ref{theorem_local_property} as $\epsilon_{x,x'} = \sup_{t \in \mathcal{Y}}\ell_{x,x'}(t)$ and shows that DP is a local property. Here the term “local” is used as common in real analysis, referring to features of a function, that are determined by its behavior in only a small neighborhood (in the case of $\mathcal{\ell}_{x,x'}$ in a neighborhood around its argmax).   

\begin{figure}[H] 
\centering
\includegraphics[width=1\linewidth,height=180pt]{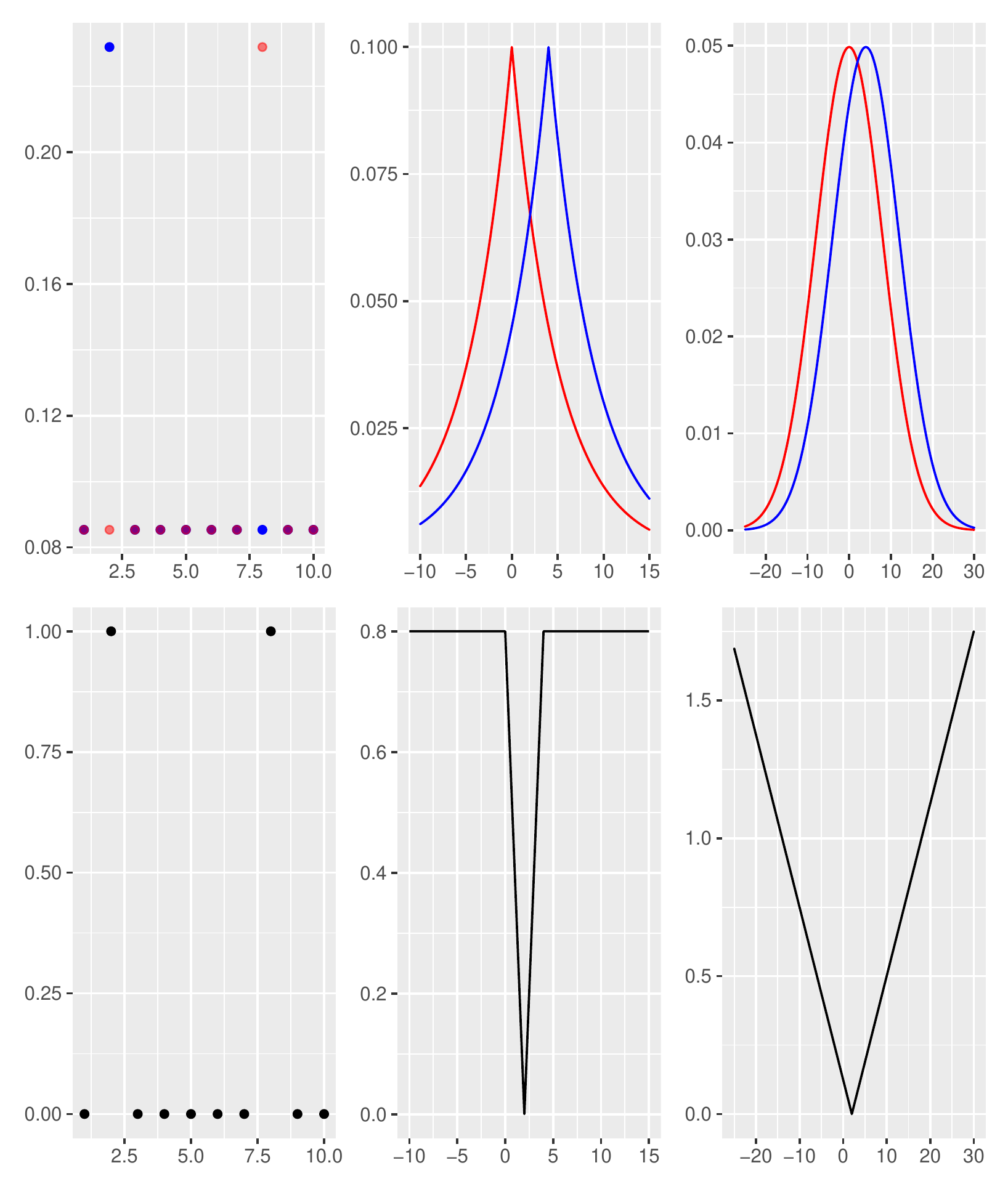}\\
\caption{The top row depicts the densities $f_x \sim A(x), f_{x'} \sim A(x')$ for two neighboring databases $x,x'$ and algorithm $A$ chosen (from left to right) as Randomized Response, the Laplace Mechanism and Gaussian Mechanism. The bottom row captures the corresponding loss functions $\ell_{x,x'}$ from \eqref{Eq_def_loss_function}.} \label{standard_examples}
\end{figure}

Figure \ref{standard_examples} provides an illustration of the loss function for some standard examples of randomized algorithms (see e.g. \cite{Dwork2014, Kairouz2016}). The plots help discern the amount of privacy leakage and where it occurs. For example, we observe that for Randomized Response (left) only two outputs elicit any privacy leakage at all, while the maximum loss associated with the  Laplace Mechanism (middle panel) is assumed everywhere, except for the area enclosed by the density modes.
For the Gaussian Mechanism (right panel) no single $t$ exists that maximizes the loss. Instead, $\ell_{x,x'}(t)$ tends to infinity for growing $|t|$, which implies decreasing privacy for tail events. The unbounded loss function for $\vert t \vert \to \infty $ shows that the Gaussian Mechanism does not satisfy pure DP.

In the next section, we develop statistical methods based on Theorem \ref{theorem_local_property}. Before doing so, we want to point out the possibilities and limitations of this approach. Theorem \ref{theorem_local_property} presupposes that an algorithm under consideration must be either discrete or continuous. One counterexample from the related literature is a flawed version of the Sparse Vector Technique (Algorithm 3 in \cite{Lyu2017}), which is neither fully continuous nor discrete and therefore lies outside the scope of our methods. Still, we want to emphasize that algorithms usually considered in the validation literature fall into either category (in \cite{DP-Sniper} all except for SVT3, SVT34Parallel and NumericalSVT, which are all variations of the above Sparse Vector Technique).

The key advantage of dividing algorithms into continuous and discrete ones is that we can tailor estimation methods to each case. This notably helps us to handle the tricky case of continuous algorithms. More precisely, continuous algorithms will assume any value on a continuum (e.g. an interval) and therefore the ensuing output space is infinitely large. To appreciate the practical effects of this, consider a discretization of the output space: Suppose we discretize the unit interval $\mathcal{Y} = [0,1]$ into $1000$ equally spaced points $\mathcal{Y}^{discr}:= \{1/1000,...,999/1000, 1\}$. This discretization may seem modest in terms of precision, but it already yields an output space of $1000$ distinct elements. 

Why is this a problem? As the grid gets finer, the output probability of any $t \in \mathcal{Y}^{discr}$ decreases and the sampling effort to approximate the probability soars (at least for standard estimators like the empirical measure used in \cite{StatDP} and \cite{DP-Sniper}). It is thus hard to assess DP on small events, which however is key for general, continuous algorithms.

To resolve this issue, we turn to the theory of kernel density estimation: Instead of relying on the all-or-nothing information “$A(x) = t $” vs “$A(x) \neq t $” (as the empirical measure does), KDE draws on the more gradual information “$A(x)$ is near $t$”. While sampling a certain output $t$ in the continuous case may be unlikely (impossible even from a theoretical perspective), drawing a sample with some values close to $t$ is highly probable. This implies that KDE can provide reliable estimates even of small probabilities, which do not depend on the grid size of a discretization and only on the smoothness of the underlying density (see Section \ref{Subsec_2_B}).\\

\noindent We briefly summarize the \textbf{key insights of this section}:

Instead of examining large and complex sets in order to quantify $\epsilon_{x,x'}$, Theorem \ref{theorem_local_property} shows that it suffices to consider single output values $t \in \mathcal{Y}$. In fact, larger events $E$ potentially dilute the observed privacy violation and lead to an underestimation of privacy leakage.
Numerically, the task of maximizing  $L_{x,x'}$ (a function with sets as arguments), is much more difficult than to maximize $\ell_{x,x'}$ (which has arguments in $\mathbb{R}^d$ or $\mathbb{N}$), where standard solutions exist (see \cite{forst2010}). 
Finally, the loss function $\ell_{x,x'} $ is far more amenable to interpretation than $L_{x,x'}$. In fact, $\ell_{x,x'}$ can be plotted and thus problematic areas with respect to privacy can be easily displayed and understood (e.g., we see at one glance, that for the Gaussian Mechanism, which only satisfies approximate DP, the problem lies in extreme values of $t$; see Figure 2, right).

\noindent We conclude this section with a non-trivial example, where we utilize the loss function to derive the privacy parameter $\epsilon$.

\begin{example} \label{example_1} 
We consider a database $x$ containing the information of only one individual ($m = 1$). Assuming that said individual's data is a vector $
    v = (v_1, \cdots, v_k) \; \in \; [0,1]^k,
$ i.e. $\mathcal{D} = [0,1]^k$, we can identify our database as $x = v$. It is our intention to publish the maximum entry of $v$ in a differentially private manner. We can do this by employing a version of the \textit{Noisy Max} algorithm (Algorithm 7 in \cite{StatDP}) where we add independent Laplace noise $L_i \sim Lap(0, \frac{1}{\lambda})$ to each component $v_i$ and publish the maximum $\max_i (v_i+L_i)$. We demonstrate how $\ell_{x,x'}$ can be used to determine the privacy parameter $\epsilon$ of this algorithm. \\
On the one hand, releasing a noisy component $v_i + L_i$ by itself satisfies $\lambda$-DP by virtue of the Laplace Mechanism. The maximum can then be understood as a function over the vector of noisy components and the composition theorem of DP yields $ k \lambda$ as an upper bound of $\epsilon$. 
On the other hand, define $F_i$ as the distribution function of $v_i+L_i$ and $f_i = F_i'$ as the corresponding density. Then the density $f_v$ of the random variable $\max_i (v_i+L_i)$ is of the form
$$
f_v(t) = \Big(\sum_{i=1}^k \frac{f_i(t)}{F_i(t)}\Big) \Big( \prod_{i=1}^k F_i(t) \Big).
$$
In the case where $v_1=...=v_k$, this can be simplified to $f_v(t)= k f_1(t) [F_1(t)]^{k-1}$. Using this formula, it is a straightforward calculation to show that for $v=(0,...,0)$, $w=(1,...,1)$ and  sufficiently large $t \in \mathbb{R}$
$$
 \ell_{v,w}(t) = |\ln(f_v(t))-\ln(f_w(t))| = k \lambda.
$$
Theorem \ref{theorem_local_property}  especially implies that $k \lambda$ is also a lower bound of $\epsilon$ and thus the equality $\epsilon = k \lambda $ holds.
\end{example}

$ $\\[-6ex]

\section{Quantifying the Maximum Privacy Violation} \label{Sec_4}
In this section, we proceed to the statistical aspects of our discussion.
According to Theorem \ref{theorem_local_property} the data-specific privacy 
violation $\epsilon_{x,x'}$ defined in \eqref{lower_bound} can be attained by maximizing the loss function $\ell_{x,x'}$ defined in \eqref{Eq_def_loss_function}. 
We devise an estimator $\hat \epsilon_{x,x'}$ for $\epsilon_{x,x'}$, 
by maximizing an empirical version $\hat \ell_{x,x'}$ of the loss function, specified in Section \ref{subsec_41}.
In Proposition \ref{proposition_1}, we demonstrate mathematically that such estimators are consistent with fast convergence rates. 
Besides estimation, we consider confidence intervals for the pointwise privacy loss $\ell_{x,x'}(t)$ in Section \ref{subsec_42}. If applied to a $t^*$ close to the argmax of $\ell_{x,x'}$, these can be used to statistically locate $\epsilon_{x,x'}\approx \ell_{x,x'}(t^*)$.

Next recall that the global privacy parameter $\epsilon$ as well as the data-centric privacy level $\epsilon_x$, defined in \eqref{Eq_max_violation} and \eqref{Eq_def_epsx} respectively, can
be attained by maximizing $\epsilon_{x,x'}$ over a (sub)space of databases. It therefore makes sense
to approximate them (from below) by a finite maximum, s.t. for instance
\begin{equation}\label{Eq_max_right}
\epsilon \approx  \max (\epsilon_{x_1, x_1'},..., \epsilon_{x_B, x_B'}),
\end{equation}
where $(x_1, x_1'),...,(x_B, x_B')$ are $B$ pairs of adjacent databases (approximating $\epsilon_x$ works by setting $x=x_1=...=x_B$). If the databases are chosen appropriately, the maximum on the right side of \eqref{Eq_max_right} comes arbitrarily close to $\epsilon$. Prior work suggests that oftentimes simple heuristics already yield databases that point to the global privacy parameter $\epsilon$ \cite{StatDP}. 
Furthermore, the structure of the data space $\mathcal{D}$ can naturally motivate search patterns (typically choosing $x_b$ and $x_b'$ to be “far apart” in some sense).

We use the approximation in \eqref{Eq_max_right}, combined with our estimators for the data-specific privacy violations, for the statistical inference of the parameters $\epsilon$ and $\epsilon_x$.
We integrate these methods into the MPL algorithm presented in Section \ref{subsec_43} and demonstrate that its output $[LB, \infty)$ is a one-sided, asymptotic confidence interval (Theorem \ref{theorem_3}).

\subsection{Estimating data-specific privacy violations} \label{subsec_41}

We now consider the problem of estimating the data-specific privacy violation $\epsilon_{x,x'}$ for two  adjacent databases $x,x'$ defined in \eqref{lower_bound}. According to Theorem \ref{theorem_local_property} we can express $\epsilon_{x,x'}$ as the maximum of the loss function $ \ell_{x,x'}$, i.e.
$$
\epsilon_{x,x'}=\sup_{t \in \mathcal{Y}} \ell_{x,x'}(t),
$$ 
where $ \ell_{x,x'}$ is defined in \eqref{Eq_def_loss_function}.
It stands to reason to first estimate the privacy loss $\ell_{x,x'}$ by an empirical version $\hat \ell_{x,x'}$, which is then maximized to obtain an estimate for $\epsilon_{x,x'}$. Suppose that $A$ is either discrete or continuous, s.t. a realization of $A(x)$ has density $f_x$. By running that algorithm $n$ times on databases $x$ and $x'$ respectively, we can generate two independent samples of i.i.d observations $X_1,...,X_n \sim f_x$ and $Y_1,...,Y_n \sim f_{x'}$. Recalling the definition of the loss function in \eqref{Eq_def_loss_function}, we can naturally define the \textit{empirical loss function} as
\begin{align} \label{Eq_def_emp_loss}
\hat{\ell}_{x,x'}(t) := \big| \ln (\hat f_x(t)) - \ln (\hat f_{x'}(t)) \big|,
\end{align}
where $\hat f_x, \hat f_{x'}$ are  density estimators for $f_x, f_{x'}$. In the case of continuous densities, we can obtain such estimators via the TKDE algorithm (see Section \ref{Subsec_2_B}). For discrete densities, we can use a truncated version of the relative frequency estimator, which is described in the TDDE (truncated discrete density estimator) algorithm and mathematically defined as follows:
$$
\hat f_x (t) := \frac{|\{X_i: X_i=t\}|}{n} \lor \tau.
$$
As in the TKDE algorithm “$\lor$” denotes the maximum and $\tau>0$ a floor to avoid instabilities due to small probabilities. The floor can be chosen smaller if $n$ is larger and the density estimate more accurate. We formalize this in the following assumption for discrete algorithms: \\

\begin{itemize}
    \item[(D)] The parameter $\tau$ is adapted to $n$ and satisfies $\tau= \mathcal{O}(\ln(n)/\sqrt{n})$. \\
\end{itemize}

\begin{algorithm}
	\caption{Truncated discrete density estimator} \label{TDDE}
	\small
	\algorithmicrequire \, $X=(X_1,...,X_n)$: data sample,   $t$: evaluation point, $\tau$: floor \\
	\algorithmicensure \, $\hat{f}(t)$: density estimate at point $t$
	\begin{algorithmic}[1]
	\Function{TDDE}{$ X,t, \tau$}\\
	    $out := 0$
		\For {$i=1,2,\ldots,n$}
			\If {$ X_i = t$} 
				\State $out = out+ 1$
			\EndIf
		\EndFor \\
    
	$out = out/n$ \\
		\Return $out \lor \tau$
	\EndFunction
	\end{algorithmic} 
\end{algorithm}

In principle, we could now approximate  $\epsilon_{x,x'}$ by maximizing the empirical loss $\hat \ell_{x,x'}$. Yet for algorithms with large output spaces (in particular continuous algorithms) $\hat \ell_{x,x'}$ can yield unreliable estimates for extreme values of $t$, where (almost) no observations are sampled. 
We therefore restrict maximization to a closed, bounded set $C \subset \mathcal{Y}$, usually an interval (or hypercube in the multivariate case). Notice that 
\begin{equation} \label{Eq_truncated_level}
    \epsilon_{x,x',C} := \sup_{t \in C} \ell_{x,x'}(t)  \approx \sup_{t \in \mathcal{Y}} \ell_{x,x'}(t) = \epsilon_{x,x'}
\end{equation}
in the sense that the difference between $\epsilon_{x,x',C}$ and $\epsilon_{x,x'}$ can be made arbitrarily small for sufficiently large $C$.
For most standard algorithms even strict equality holds for some fixed $C$ (as is the case for all algorithms investigated in Section \ref{Sec_5}). 
This is in particular true for discrete algorithms with finite range, where we can always choose $C = \mathcal{Y}$. 

We now state two regularity conditions that pertain to continuous algorithms and guarantee reliable inference: \\

\begin{itemize}
    \item[(C1)] There exists a constant  $\beta \in (0,1]$, such that for all $x$ the density $f_{x}$ corresponding to $A(x)$ is $\beta$-Hölder continuous.\\
    \item[(C2)] For any $x, x'$ and any sequence $(t_n)_{n \in \mathbb{N}}$ in $C$, which satisfies 
    $$
    \lim_{n \to \infty} \ell_{x,x'}(t_n) = \sup_{t \in C} \ell_{x,x'}(t),
    $$
    it holds that $(t_n)_{n \in \mathbb{N}}$ has a limit point in $\argmax_{t \in C }\ell_{x,x'}(t)$. \\
\end{itemize} 

We briefly comment on these assumptions:
Condition (C1) demands that our algorithm is not only continuous in the sense that it has probability densities everywhere, but that these additionally satisfy a weak regularity condition of $\beta$-smoothness (see Section \ref{Subsec_2_B}).  
This guarantees reliable kernel density estimators and thus a good approximation of $\ell_{x,x'}$ by $\hat \ell_{x,x'}$.
Condition (C2) is a technical requirement that appears more complicated than it is: It prohibits the maximum privacy violation (of $A$ on $C$) from occurring in locations where both densities are $0$, thus excluding pathological cases. Many continuous algorithms satisfy both of these conditions (among them all those discussed in this paper).

We now define the location $\hat t$ of maximum privacy violation:
\begin{align} \label{Eq_def_hat_t}
\hat{t} \in  \argmax_{t \in C}   \hat{\ell}_{x,x'}(t) .
\end{align}
In the following we demonstrate that the maximum of the empirical loss function, i.e.
\begin{equation} \label{Eq_def_hat_eps}
    \hat \epsilon_{x,x'} := \hat \ell_{x,x'}(\hat t)  
\end{equation}
is close to the maximum of the true loss function. 

\begin{figure}
\centering
\includegraphics[width=0.8\linewidth,height=140pt]{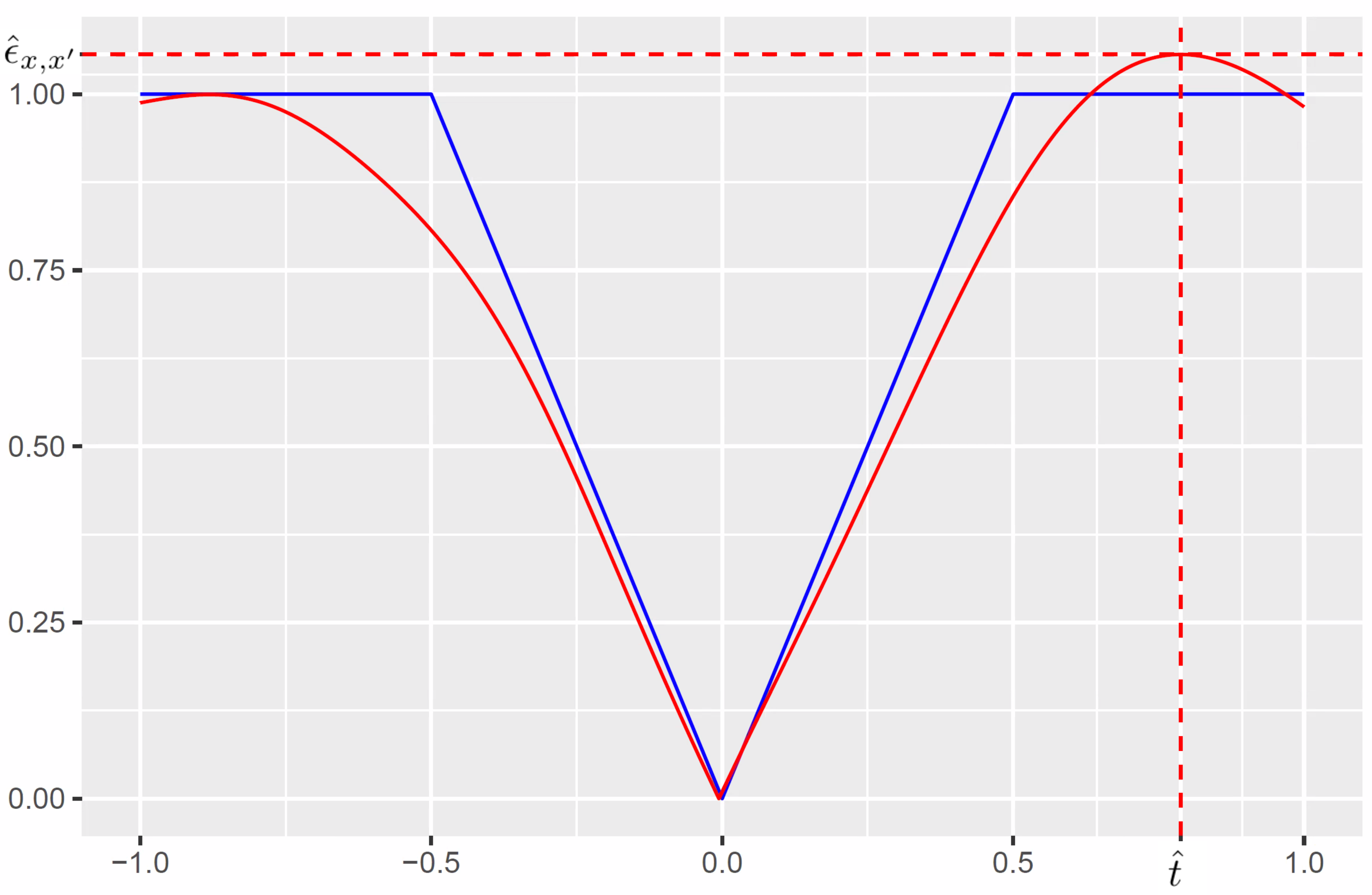}\\
\caption{Loss function $\ell_{x,x'}$ (blue) and empirical loss $\hat \ell_{x,x'}$ (red) for the Laplace algorithm. The vertical line indicates the location of the argmax $\hat t$ and the horizontal line the maximum $\hat \epsilon_{x,x'}$ of the empirical loss function.}
\end{figure}
To derive asymptotic convergence rates in the continuous case, the bandwidths $h$ and $h'$ of the truncated kernel density estimators $\hat{f}_x$ and $\hat{f}_{x'}$ in \eqref{Eq_def_emp_loss} have to be chosen appropriately.
In addition, the floor $\tau $ must not  be smaller than the precision level of the density estimators (see Section \ref{Subsec_2_B}). We specify the proper choice of parameters in the following condition: \\

\begin{itemize}
   \item[(C3)] The parameters $h, h'$ and $\tau$ are adapted to $n$ and satisfy
    $$
    h, h'= \mathcal{O}\big(n^{-\frac{1}{2\beta+d}}\big), \,\,\quad \tau = \mathcal{O}\big(n^{-\frac{\beta}{2\beta+d}} \ln(n)\big).
    $$ 
\end{itemize} 

\begin{proposition} \label{proposition_1}
Suppose that $C$ is a closed, bounded set and $\epsilon_{x,x',C}\in (0, \infty)$. If $A$ is a discrete algorithm and condition $(D)$ is satisfied, it follows that
\begin{align*}
& |\hat \epsilon_{x,x'} - \epsilon_{x,x',C}| = \mathcal{O}_P(n^{-1/2})\\
\textnormal{and} \quad & |\ell_{x,x'}(\hat t) - \epsilon_{x,x',C}| = \mathcal{O}_P(n^{-1/2}).
\end{align*}
If $A$ is a continuous algorithm such that conditions $(C1)-(C3)$ are satisfied, it follows that
\begin{align*}
&
|\hat \epsilon_{x,x'}  - \epsilon_{x,x',C}| =  \mathcal{O}_P \Big( \sqrt{\ln(n)} n^{-\frac{\beta}{2\beta+d}}\Big)\\
\textnormal{and} \quad &
|\ell_{x,x'}(\hat t) - \epsilon_{x,x',C}| =  \mathcal{O}_P \Big( \sqrt{\ln(n)} n^{-\frac{\beta}{2\beta+d}}\Big).
\end{align*}
Furthermore, if $\epsilon_{x,x',C} \in \{0, \infty\}$ it holds that 
$$
\hat \epsilon_{x,x'} \to_P \epsilon_{x,x',C} 
$$  where “$\to_P$” denotes convergence in probability (see Appendix A for a definition).
\end{proposition}

The first identity for both the discrete and continuous case in Proposition \ref{proposition_1} suggests that the maximum privacy violation for $x, x'$ is approximated by its empirical counterpart at the same rate as the densities $f_x, f_x'$ by their  estimators, which again is different in both settings. This rate -specifically in the continuous case- should not be taken for granted: Admittedly, if the two continuous densities $f_x, f_{x'}$ are bounded away from $0$ on $C$, it is not difficult to show that
$$
\sup_{t \in C}|\hat \ell_{x,x'}(t)- \ell_{x,x'}(t)| = \mathcal{O}_P\Big(\sqrt{\ln(n)} n^{-\frac{\beta}{2\beta+d}}\Big),
$$
which implies the Proposition. However, if the densities are not bounded away from $0$, it may not be true that $\ell_{x,x'}$ is uniformly approximated by $\hat \ell_{x,x'}$. Still, the approximation of the maxima  holds and is not slowed down in this case (even though the mathematical proof gets substantially more involved).  

The second identity (for both cases) states that $\hat t$ is close to the argmax of $\ell_{x,x'}$ in the sense that the true loss function evaluated at $\hat t$ is close to its maximum on $C$. This fact will be used in the two subsequent sections, where we argue that a confidence interval for $\ell_{x,x'}(\hat t)$ automatically contains $\epsilon_{x,x',C}$.

We conclude this section by stating the DPL algorithm (Algorithm 3) which, given $x$ and $x'$, calculates the maximum empirical privacy loss, as well as $\hat{t} $. In DPL, the binary variable \textit{$discr$} indicates whether a discrete (1) or continuous (0) setting is on hand and the set $C$ encloses the area of interest.

\begin{algorithm}
	\caption{Data-specific privacy loss} 
	\small
	\algorithmicrequire \; \parbox[t]{\dimexpr\linewidth-\algorithmicindent}{
           neighboring databases $x$ and $x'$, closed and bounded set $C$, \\ sample size $n$, specification variable $discr$
        } 
	\algorithmicensure \, estimated loss $\hat{\epsilon}_{x,x'}$, location of loss $\hat{t}$
	\begin{algorithmic}[1]
	\Function{DPL}{$x,x',n,C,discr$}
	\State Generate $X = (X_1, \cdots, X_n)$ with $X_i \sim A(x)$
	\State Generate $Y = (Y_1, \cdots, Y_n)$ with $Y_i \sim A(x')$
	\State Set $\tau$ in accordance with (D) if $discr=1$
	\State Set $h,h'$ and $\tau$ in accordance with (C3) if $discr=0$
	\State Choose appropriate kernel $K$
    	\If {$discr = 1$} 
		    \State  $\hat f_{x} (\cdot)=\textnormal{TDDE}(X,\cdot, \tau)$
	        \State $\hat f_{x'} (\cdot)=\textnormal{TDDE}(Y,\cdot, \tau)$
		\Else 
		    \State $\hat{f}_{x}(\cdot) = \textnormal{TKDE}(X, \cdot,h,K,\tau)$
	        \State $\hat{f}_{x'} (\cdot) = \textnormal{TKDE}(Y, \cdot,h',K, \tau)$
		\EndIf
		\State $\hat{\ell}_{x,x'}(\cdot) = \vert \ln(\hat{f}_{x}(\cdot)) - \ln(\hat f_{x'} (\cdot)) \vert $
    	 \State $\hat{t} = \argmax \{ \hat{\ell}_{x,x'}(t): t \in C\}$ \State  $\hat{\epsilon}_{x,x'} = \hat{\ell}_{x,x'}(\hat{t})$
    	 \State \Return $(\hat{t},\hat{\epsilon}_{x,x'})$
    \EndFunction
	\end{algorithmic} 
\end{algorithm}

\subsection{Statistical bounds for pointwise privacy loss}\label{subsec_42}
In the previous section, we have considered the problem of estimating data-specific privacy violations. We now move to the related topic of statistical inference in the sense of Section \ref{Subsec_2_A}: Finding a confidence interval for $\epsilon_{x,x',C}$. \\
More precisely, we show in this section how to construct an asymptotic confidence interval for the pointwise privacy loss $\ell_{x,x'}(t)$ for an arbitrary $t \in C$, which we apply later to the choice $t= \hat t$ (recall that according to Proposition \ref{proposition_1} we have  $\ell_{x,x'}(\hat t)\approx \epsilon_{x,x',C}$).

Suppose that $\ell_{x,x'}(t) \in (0,\infty)$. In this situation it can be shown by asymptotic normality of the density estimators and the delta method (see \cite{vandervaart1998}), that for all $t \in \mathbb{R}$
\begin{equation} \label{Eq_weak_convergence}
  \lim_{n \to \infty} \mathbb{P}\Big( \frac{c_n}{\sigma} (\hat \ell_{x,x'}(t)-\ell_{x,x'}(t)) \le t \Big) = \Phi(t).
\end{equation}
Here $\Phi(\cdot)$ is the distribution function of a standard normal random variable and $c_n= \sqrt{n}$ if the algorithm $A$ is discrete and $c_n=\sqrt{n h^d}$ if it  is continuous. In the latter case $h$ denotes the bandwidth of both $\hat f_x, \hat f_{x'}$ and is assumed to be adapted to the sample size $n$ as $h= \mathcal{O}(n^{-\frac{1}{2\beta+d}-\gamma})$ for some $\gamma>0$. This bandwidth is smaller than the one suggested in (C3) and leads to a slower uniform convergence of the corresponding density estimators (see Section \ref{Subsec_2_B}, \eqref{Eq_KDE_rate_2}). Such a bandwidth choice, which makes the variance of the density estimator larger than its bias, is referred to as “undersmoothing”. Undersmoothing is a standard tool in the statistical analysis of continuous densities, where  the two tasks of estimation and inference require different degrees of smoothing (see \cite{Horowitz2001} p.3999).

The variance $\sigma^2$ on the right side of \eqref{Eq_weak_convergence} can be expressed as follows:
$$
\sigma^2 := \begin{cases}
\frac{1}{f_x(t)}+\frac{1}{f_{x'}(t)}-2, \quad\quad\quad\quad\quad A \,\, \textnormal{discrete}
\\
\int K^2(s) \, ds \, \Big(\frac{1}{f_x(t)}+\frac{1}{f_{x'}(t)} \Big), \quad A \,\, \textnormal{continuous}.
\end{cases}
$$

Note that $\sigma^2$ is well-defined in both cases (in particular in the discrete case $1/f_x(t), 1/f_{x'}(t)>1$, s.t. the variance is indeed positive). Also notice that $\sigma^2$ is unknown, but easy to estimate in practice, replacing the true densities by their estimators $\hat f_x, \hat f_{x'}$, which yields
$$
\hat \sigma^2 := \begin{cases}
\frac{1}{\hat f_x(t)}+\frac{1}{\hat f_{x'}(t)}-2, \quad\quad\quad\quad\quad A \,\, \textnormal{discrete}
\\
\int K^2(s) ds \Big(\frac{1}{\hat f_x(t)}+\frac{1}{\hat f_{x'}(t)} \Big), \quad A \,\, \textnormal{continuous}.
\end{cases}
$$
It is straightforward to show that $\hat \sigma^2 = \sigma^2 +o_P(1)$. We can now use this fact, together with the convergence in \eqref{Eq_weak_convergence}, to see that for any $\alpha \in (0,1)$
\begin{align}\label{Eq_LB}
1-\alpha \approx & \,\, \mathbb{P}  \Big(  \frac{c_n}{\hat \sigma} (\hat \ell_{x,x'}(t)-\ell_{x,x'}(t)) \le \Phi^{-1}(1-\alpha)  \Big)\\
=&\,\, \mathbb{P} \Big( \hat \ell_{x,x'}(t)+\frac{\Phi^{-1}(\alpha) \hat \sigma }{c_n} \le \ell_{x,x'}(t) \Big). \nonumber
\end{align}
Here $\Phi^{-1}$ denotes the quantile function of the standard normal distribution and we have used the identity $\Phi^{-1}(1-\alpha) = -\Phi^{-1}(\alpha)$.
The approximation of $1-\alpha$ by the probability gets more accurate as the sample size $n$ increases and we see that 
$$
\hat I_\alpha := [\hat \ell_{x,x'}(t)+ \hat \sigma c_n^{-1}\Phi^{-1}(\alpha), \infty) 
$$
is an asymptotic confidence interval for $\ell_{x,x'}(t)$ (in the sense of Section \ref{Subsec_2_A}).

\subsection{A statistical procedure for the maximum privacy violation} \label{subsec_43}
Recall the definition of $\epsilon_{x,x',C}$ in \eqref{Eq_truncated_level}. In this section we construct the algorithm called MPL (Maximum Privacy Loss) whose output $LB$ lower bounds  the maximum of $\epsilon_{x_1,x_1',C},...,\epsilon_{x_B,x_B',C}$ with prescribed probability $1-\alpha$. The choice of $\alpha$ is determined by the user but, guided by common practice in hypothesis testing, we recommend $\alpha \in \{0.1, 0.05, 0.01\}$. By construction the inequality $$
\max\{\epsilon_{x_1,x_1'},...,\epsilon_{x_B,x_B'} \} \ge \max\{\epsilon_{x_1,x_1',C},...,\epsilon_{x_B,x_B',C} \}
$$
holds and both sides are arbitrarily close for large enough $C$. Hence, $LB$ will also constitute a tight lower bound for the maximum on the left and thus of the privacy parameter $\epsilon$ (see \eqref{Eq_max_right}). An outline of MPL is given in Algorithm 4.

We now study the structure of the MPL algorithm, which 
calculates $LB$ for a given set 
$$
\mathcal{X} = \{ (x_1, x_1'),...,(x_B, x_B') \} 
$$
of $B$ adjacent pairs and is composed of two parts.
The first part of the algorithm is dedicated to finding the pair of databases $(x_{max},x_{max}') \in \mathcal{X}$ along with the corresponding location $\hat t_{max}$ that maximize the empirical privacy violation. For that purpose, MPL runs the DPL algorithm for each pair $(x_b, x_b')$  to approximate the data-specific privacy violation $\epsilon_{x_b, x_b'}$ by an estimate $\hat \epsilon_{x_b, x_b'}$. Based on the empirical violations $\hat \epsilon_{x_1, x_1'},..., \hat \epsilon_{x_B, x_B'}$, the pair of databases $(x_{max},x_{max}')$ with the highest privacy loss is chosen. The location where the empirical privacy loss $\hat \ell_{x_{max},x_{max}'}$ is maximized is called $\hat t_{max}$ (which is an output of $\textnormal{DPL}$ run on $(x_{max},x'_{max})$). Structurally, this part of the algorithm resembles counterexample generation \cite{StatDP} and the tuple $(\hat \epsilon_{x_{max}, x'_{max}}, x_{max}, x'_{max}, \hat{t}_{max})$ already yields useful information concerning the location and magnitude of the maximum privacy violation.

The second part of the MPL algorithm is designed to establish a confidence region for the privacy loss at  $(x_{max}, x'_{max}, \hat{t}_{max})$. Notice that by construction $\ell_{x_{max},x_{max}'}(\hat t_{max})\approx \epsilon_{x_{max}, x_{max}'}$ holds (see Proposition \ref{proposition_1}) and that therefore said confidence region captures the maximum privacy violation. The methods for deriving $LB$ are borrowed from Section \ref{subsec_42} and are performed independently from the first part of the algorithm. MPL creates two fresh samples $X_1^*,...,X_N^* \sim A(x_{max})$ and $Y_1^*,...,Y_N^* \sim A(x'_{max})$ with sample size $N>n$. These are used to approximate the loss  $\ell_{x_{max},x_{max}'}(\hat t_{max})$ by its empirical version $\hat  \ell_{x_{max},x_{max}'}^*(\hat t_{max})$. The density estimators $\hat{f}^*_x, \hat{f}^*_{x'}$ underlying this empirical loss function are constructed with parameters $h_{max}$ and $\tau$ tailored to the construction of confidence intervals. This choice is expressed in the following condition: \\

\begin{itemize}
    \item[(C4)] Let $\nu \ge 0$. With  $N=\mathcal{O}(n^{1+\nu})$ and $\gamma>\nu/((1+\nu)6)$ we choose
    $
    h_{max}  = \mathcal{O}(N^{-\frac{1}{2\beta +d}-\gamma})
    $
    and $ \tau = o(1)$. \\
\end{itemize}

As already indicated in Section \ref{subsec_42}, bandwidths for confidence intervals have to be chosen smaller than for estimation (realized by $\gamma>0$). The trade-off between $\gamma$ and $\nu$ expresses that in the second part of the MPL algorithm, a larger sample size $N$ compared to $n$ requires more undersmoothing to control the bias. Yet, as $\nu$ is usually small  in practice (in our experiments about $0.1$), the undersmoothing requirement is rather weak. The fact that $\tau$ can decay at any rate shows that $\hat t_{max}$ (selected by truncated estimators in the first step) locates automatically in regions where the densities are not too close to $0$ and thus a second truncation by $\tau$ is not important. In applications, one could simply put $\tau = 0$ in this step.

Recalling Section \ref{subsec_42} and particularly \eqref{Eq_LB}, we can now give a confidence interval $[LB, \infty)$ for $ \epsilon_{x_{max}, x_{max}',C}$, where the statistical lower bound $LB$ is defined as follows: 
\begin{equation} \label{Eq_def_LB}
LB := \hat \ell_{x_{max},x'_{max}}^*(\hat t_{max})+\frac{\Phi^{-1}(\alpha) \hat \sigma_N}{c_N}.
\end{equation}
Here $\Phi^{-1}$ is, again, the quantile function of the standard normal distribution and $1-\alpha$ is the confidence level. The normalizing constants $c_N$ and $\hat \sigma_N$ are described in Section \ref{subsec_42}. The following theorem validates theoretically the lower bound $LB$ produced by the MPL algorithm.

\begin{theorem} \label{theorem_3} 
Suppose that $A$ is either a discrete algorithm and condition (D) is satisfied, or a continuous one such that conditions (C1)-(C4) are satisfied with regard to $A$ and the MPL algorithm. 
\begin{itemize}
    \item[i)] If 
    $$
    \epsilon_{C}^* := \max(\epsilon_{x_1,x'_1,C},...,\epsilon_{x_B,x'_B,C}) \in (0, \infty)
    $$
    it holds that
\begin{equation} \label{Eq_level}
    \lim_{n \to \infty} \mathbb{P}\Big( LB \le \epsilon_{C}^* \Big) = 1-\alpha.
\end{equation}
    \item[ii)] If $ \epsilon_{C}^* = \infty$, then $LB \to_P \infty$.  If $ \epsilon_{C}^* = 0$, then $LB \to_P 0$.
\end{itemize}
\end{theorem}
The proof of the theorem is technical and therefore deferred to the Appendix.

\begin{algorithm} 
	\caption{Maximum Privacy Loss}
	\small
	\algorithmicrequire \;\label{MPL} \parbox[t]{\dimexpr\linewidth-\algorithmicindent}{set of data pairs $\mathcal{X}$, sample sizes $n$ and $N$, region \\ of investigation $C$, specification variable $discr$, level $\alpha$} \\[0.2cm]
	\algorithmicensure \, Statistical lower bound for privacy violation $LB$
	\begin{algorithmic}[1]
	\Function{MPL}{$\mathcal{X}$, $n$, $N$, $C$, $discr$, $\alpha$}
	    \For {$b=1,\ldots,B$}
	    	\State $(\hat{t}_{x_b,x'_b},\hat{\epsilon}_{x_b,x'_b}) = \textnormal{DPL}(x_b,x'_b,n,C,discr)$
	    \EndFor 
	    \State Set $(x_{max}, x'_{max}) \in \argmax\{\hat \epsilon_{x_b, x_b'}: (x_b, x_b') \in \mathcal{X}\} $
    	\State Set $ \hat{t}_{max} := \hat{t}_{x_{max}, x'_{max}}$
	    \State Generate $X^*=(X_1^*,...,X_N^*)$ with $ X_i^* \sim A(x_{{max}})$
	    \State Generate $Y^* = (Y_1^*,...,Y_N^*)$ with $Y_i^* \sim A(x_{{max}}')$
	    \State Choose $\tau $ in accordance with (D) if $discr=1$
	    \State Choose $ h_{max},  \tau $ in accordance with (C4) if $discr=0$
	    \State Choose appropriate kernel $K$
	    \If {$discr = 1$} 
		    \State  $\hat f^*_{x_{max}} (\hat{t}_{max})=\textnormal{TDDE}(X^*,\hat{t}_{max}, \tau)$
	        \State $\hat f^*_{x'_{max}} (\hat{t}_{max})=\textnormal{TDDE}(Y^*,\hat{t}_{max}, \tau)$
		\Else 
		    \State $\hat{f}^*_{x_{max}}(\hat{t}_{max}) = \textnormal{TKDE}(X^*, \hat{t}_{max},h_{max},K,\tau)$
	        \State $\hat{f}^*_{x'_{max}} (\hat{t}_{max}) = \textnormal{TKDE}(Y^*, \hat{t}_{max},h_{max},K, \tau)$
		\EndIf
		\State $\hat{\ell}^*_{x_{max},x'_{max}}(\hat t_{max}) {\footnotesize =} \vert \ln(\hat{f}^*_{x_{max}}(\hat{t}_{max})) {\footnotesize -} \ln(\hat f^*_{x'_{max}} (\hat{t}_{max})) \vert $	
	    \State Calculate $\hat \sigma^2_N$ and $c_N$ based on $X^*, Y^*$ and $discr$
    	\State Define $LB := \hat \ell^*_{x_{max},x'_{max}}(\hat t_{max})+\frac{\Phi^{-1}(\alpha) \hat \sigma_N}{c_N}$
		\State \Return $LB$
    \EndFunction
	\end{algorithmic} 
\end{algorithm}

We conclude this section by discussing the limitations of our statistical methods with an example taken from \cite{StatDP}.  
\begin{example}
Suppose we have an algorithm $A$ that checks whether a given database $x$ matches a target database $x_0$. More precisely, we have $A(x) = 0 $ for any $x \neq x_0$ and $A(x_0) = 1$ with probability $e^{-k}$ and $A(x_0) = 0 $ with probability $1 - e^{-k}$. One can easily confirm that $A$ is not differentially private. However, for large $k$, a sampling based method such as ours could falsely identify $A$ as a constant function which trivially satisfies DP. And while $A$ is actually $(\epsilon, \delta)-DP$ for $\epsilon = 0$ and $\delta = e^{-k}$ and comes close to perfect $0-DP$, this would still amount to a misclassification of $A$.
In fact, $A$ reflects the fundamental limitations of any black box scenario where we are forced to rely solely on algorithm outputs. In order to reliably detect such intricate pathologies, one might have to ultimately access the algorithm’s source code. Here, formal verification tools (referenced in the Introduction) might be more suitable.
\end{example}

\section{Experiments} \label{Sec_5} 

In this section, we analyze the performance of our methodology by applying it to some standard algorithms 
in DP validation. We focus mainly on inference for the global privacy parameter $\epsilon$, but a subsection concerning the data-centric privacy level $\epsilon_x$ is included as well.

Our method is implemented in \texttt{R} and for kernel density estimation we use the “kdensity” package, which also provides automatic bandwidth selection.
In the following, we give a short outline of the algorithms and experiment settings before discussing our empirical findings.

\subsubsection*{\textbf{Query model}}
We briefly discuss the query model used in \cite{StatDP}. Many discrete algorithms do not operate on databases $x$ directly, but instead process query outputs $q(x)$. Thus, the search and selection of databases $x = (x(1), \cdots, x(m))$ translates into a choice of query outputs 
\begin{align*}
    q = (q_1, \cdots, q_d) = (q_1(x), \cdots, q_d(x)).
\end{align*}
Here counting queries, which check how many data points $x(i)$ in $x$ satisfy a given property, are of particular interest. A change in a single data point can affect the output of each counting query by at most $1$. Hence, query answers on neighboring databases are captured by vectors of natural numbers $q,q'$ where $q_i$ and $q'_i$ can differ by at most 1. Simple query answers that are created following patterns displayed in Table \ref{query_patterns} are sufficient to deduce the privacy parameter \cite{StatDP} and we will draw on vectors resembling these to evaluate discrete algorithms. 

\begin{table}[h]
\begin{center}
\begin{tabular}{l|c|c} 
\toprule
Pattern & Query $q$ & Query $q'$ \\ 
\midrule 
One Above & $(1,1,1,1,1,1)$ & $(2,1,1,1,1,1)$ \\
One Below & $(1,1,1,1,1,1)$ & $(0,1,1,1,1,1)$ \\
One Above Rest Below & $(1,1,1,1,1,1)$ & $(2,0,0,0,0,0)$ \\
One Below Rest Above & $(1,1,1,1,1,1)$ & $(0,2,2,2,2,2)$ \\
Half Half & $(1,1,1,1,1,1)$ & $(0,0,0,1,1,1)$ \\
All Above All Below & $(1,1,1,1,1,1)$ & $(2,2,2,2,2,2)$ \\
X Shape & $(1,1,1,0,0,0)$ & $(0,0,0,1,1,1)$ \\ 
\bottomrule
\end{tabular}
\caption{Input patterns used in \cite{StatDP}}
\label{query_patterns}
\end{center}
\end{table}

Similar to the discrete case, continuous algorithms are usually applied to aggregate statistics $S$ of the data and not to the raw data itself. We therefore consider algorithmic inputs of the form $s=S(x)$ and $s'=S(x')$, that lie in a continuous domain (in the following examples intervals and cubes).

\subsubsection*{\textbf{Algorithms}} 
We test our approach on 8 algorithms in total.
The well known \textbf{Laplace  Mechanism} (see \cite{Dwork2006}) publishes a privatized version of a real valued statistic $s \in [0,1]$ by adding centered Laplace noise $L \sim Lap( \frac{1}{\epsilon})$. This mechanism is used as a subroutine in many differentially private algorithms (e.g. the versions of Noisy Max discussed here). In the following, we consider as input statistics $s_b = 0$ and $s_b'=b/10$ for $b=1,...,10$. The set $C$ in MPL is chosen as the symmetric interval $[-1,1]$.

The \textbf{Report Noisy Max} algorithm \cite{Dwork2014} publishes the query with the largest value within a vector of noisy query answers. More precisely, the index $\argmax \{ q_i + L_i : 1 \leq i \leq d \}$ with $L_i \sim Lap( \frac{2}{\epsilon})$ is calculated and returned (see \cite{StatDP}, Algorithm 5). We implement Report Noisy Max and our procedure on vectors that entail 6 query answers and choose databases $q_b$ and $q'_b$, $b = 1, ..., 10$, that are similar to the patterns described in Table \ref{query_patterns}.

Given a query vector $q$ and a threshold $T$, the \textbf{Sparse Vector Technique (SVT)} goes through each query answer $q_i$ and reports whether said query lies above or below $T$ \cite{Dwork2014}. The maximum number of positive responses $M$ is an adjustable feature of the algorithm that forces it to abort after $M$ query answers above $T$ have been reported. We investigate 4 versions of SVT taken from \cite{Lyu2017}, which are, in accordance with the denotation in \cite{Lyu2017} and \cite{DP-Sniper}, variants SVT2 and SVT4-SVT6. We consider query vectors $q_b$ and $q'_b$, $b = 1, \cdots, 10$, with $10$ entries that are similar to the patterns in Table \ref{query_patterns}. This choice resembles the one in prior work (see \cite{StatDP, DP-Sniper}) and we do the same for the tuning parameters with $T=1$ and $M = 1$ \cite{DP-Sniper}.

\begin{figure*}
\begin{subfigure}[c]{.49\linewidth}
\centering
\caption{\textbf{Laplace Mechanism}}
\includegraphics[width=0.9\linewidth,height=140pt]{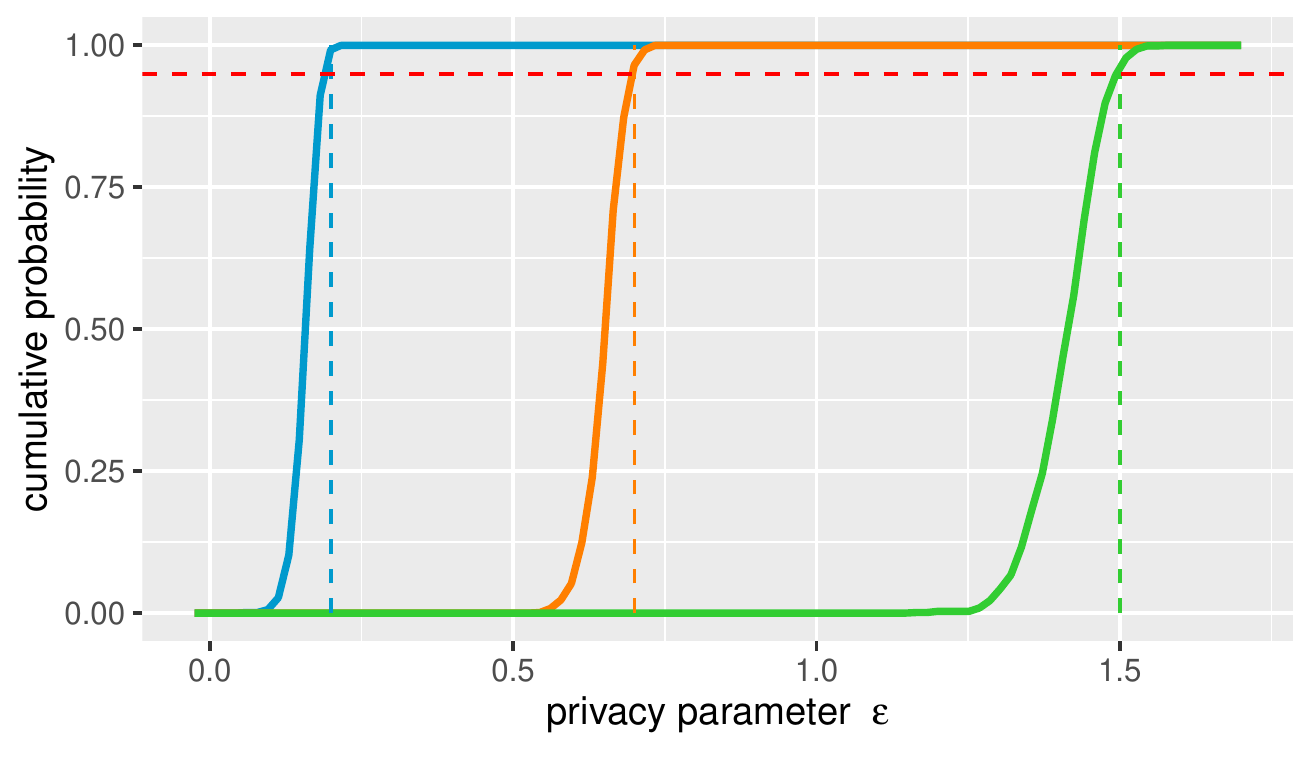}
\label{LB_noisy_plot}
\end{subfigure}
\quad
\begin{subfigure}[c]{.49\linewidth}
\centering
\caption{\textbf{Report Noisy Max}}
\includegraphics[width=0.9\linewidth,height=140pt]{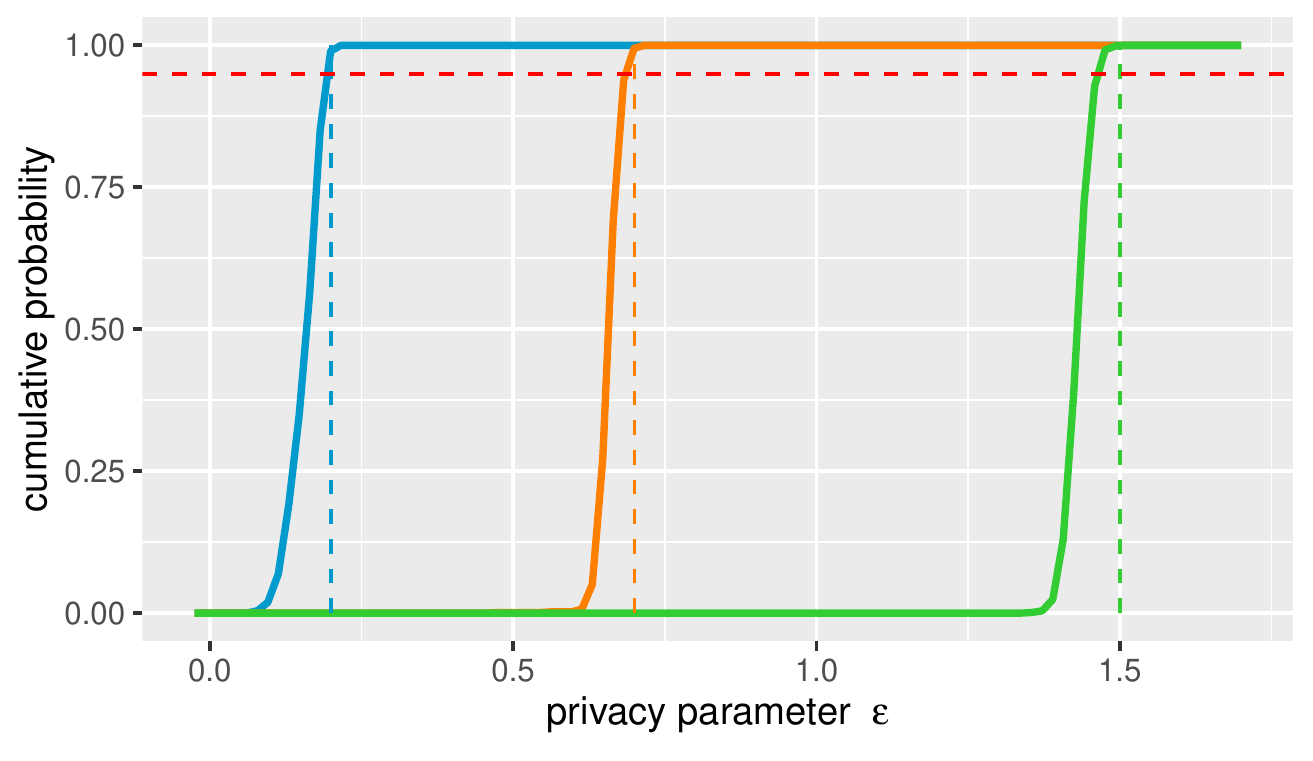}
\label{LB_svt_plot}
\end{subfigure} 

\begin{subfigure}[c]{.49\linewidth}
\centering
\caption{\textbf{Continuous Noisy Max}}
\includegraphics[width=0.9\linewidth,height=140pt]{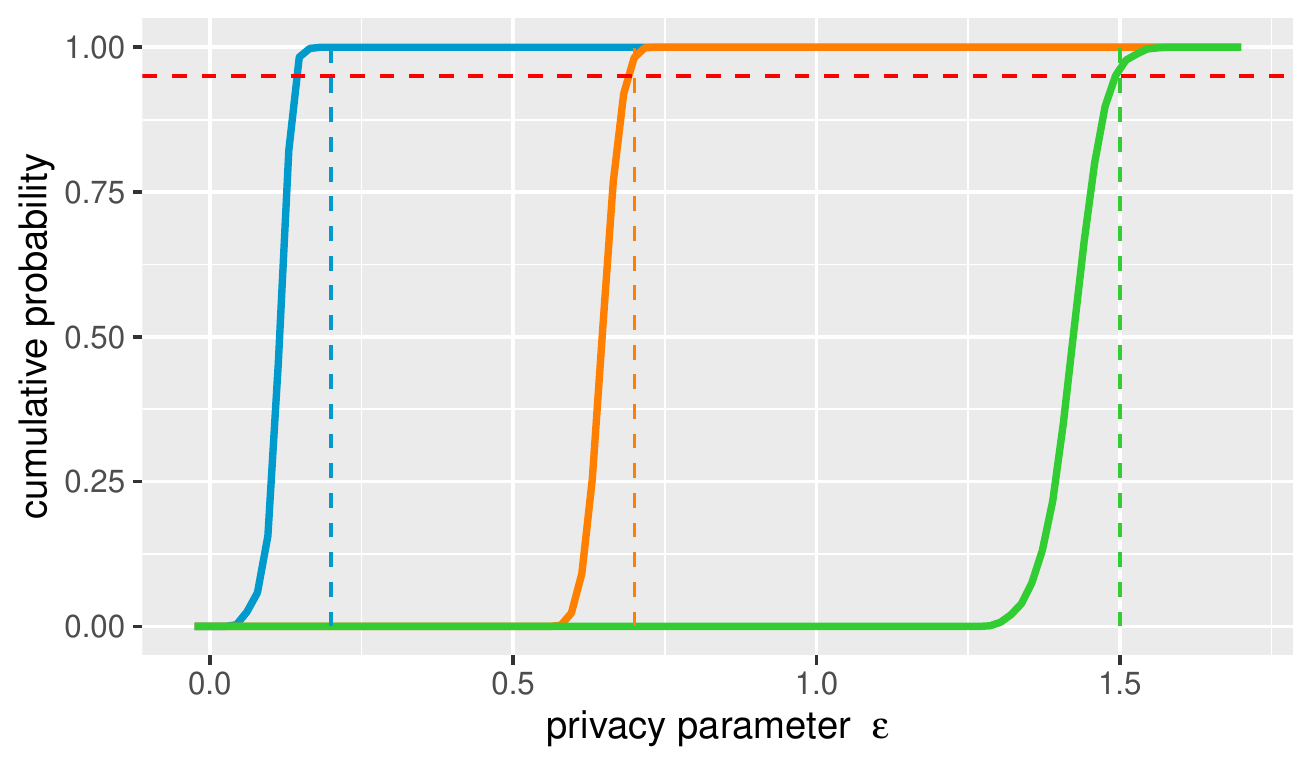}
\label{LB_max_plot}
\end{subfigure}
\quad
\begin{subfigure}[c]{.49\linewidth}
\centering
\caption{\textbf{Exponential Mechanism}}
\includegraphics[width=0.9\linewidth,height=140pt]{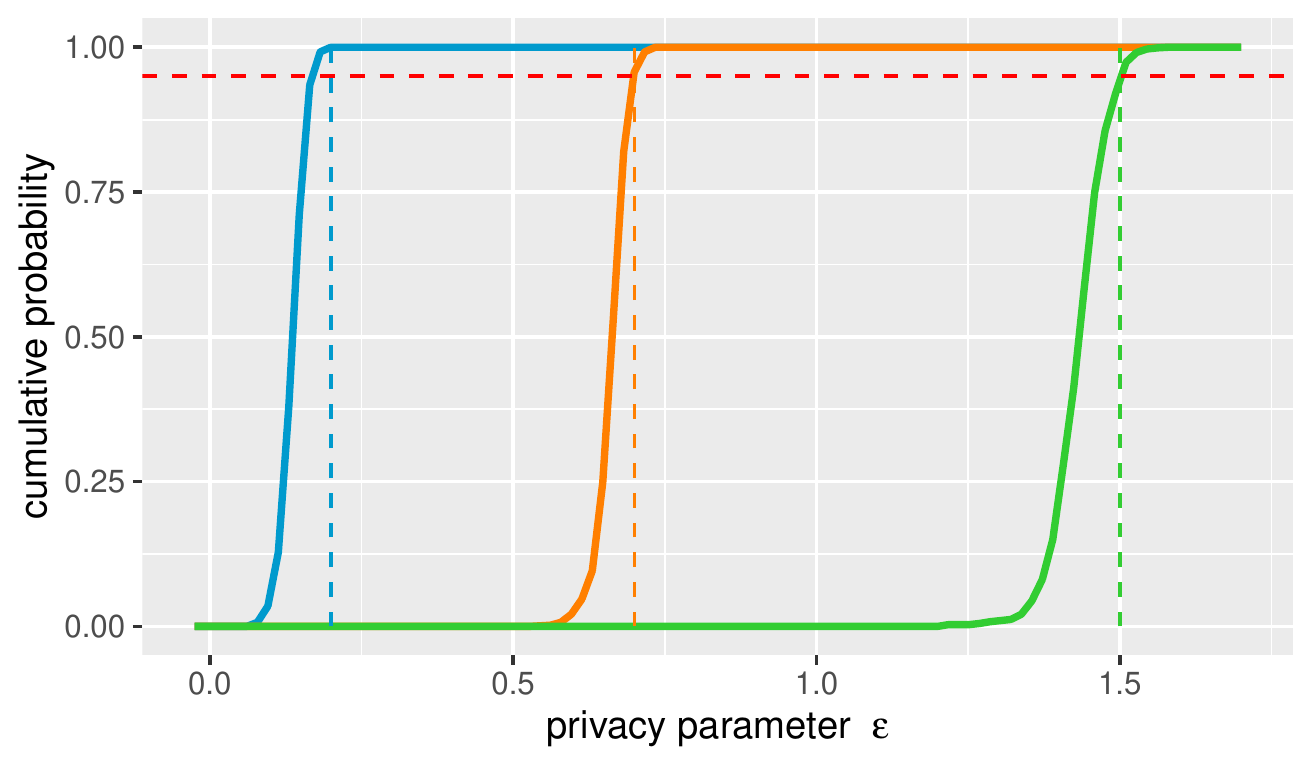}
\label{LB_exp_plot}
\end{subfigure}

\begin{subfigure}[c]{.49\linewidth}
\centering
\caption{\textbf{Sparse Vector Technique 2}}
\includegraphics[width=0.9\linewidth,height=140pt]{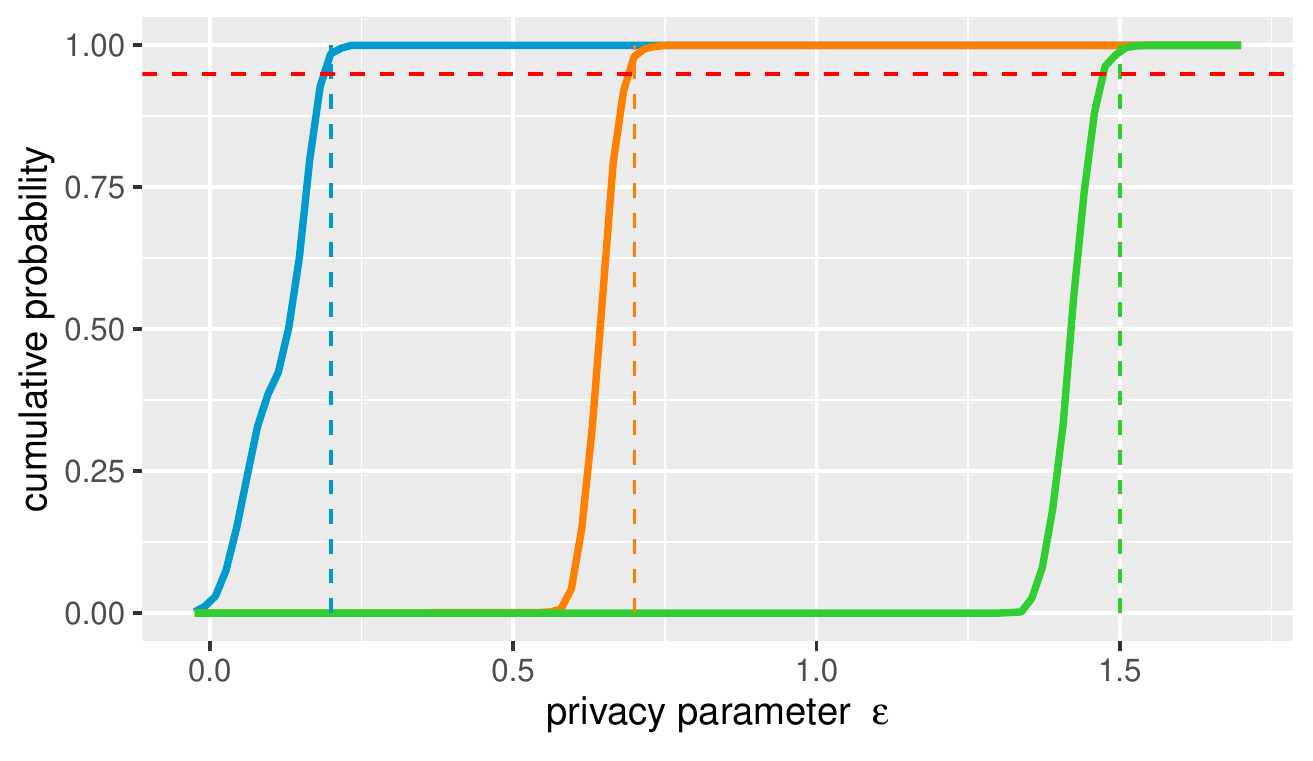}
\label{LB_max_plot}
\end{subfigure}
\quad
\begin{subfigure}[c]{.49\linewidth}
\centering
\caption{\textbf{Sparse Vector Technique 4}}
\includegraphics[width=0.9\linewidth,height=140pt]{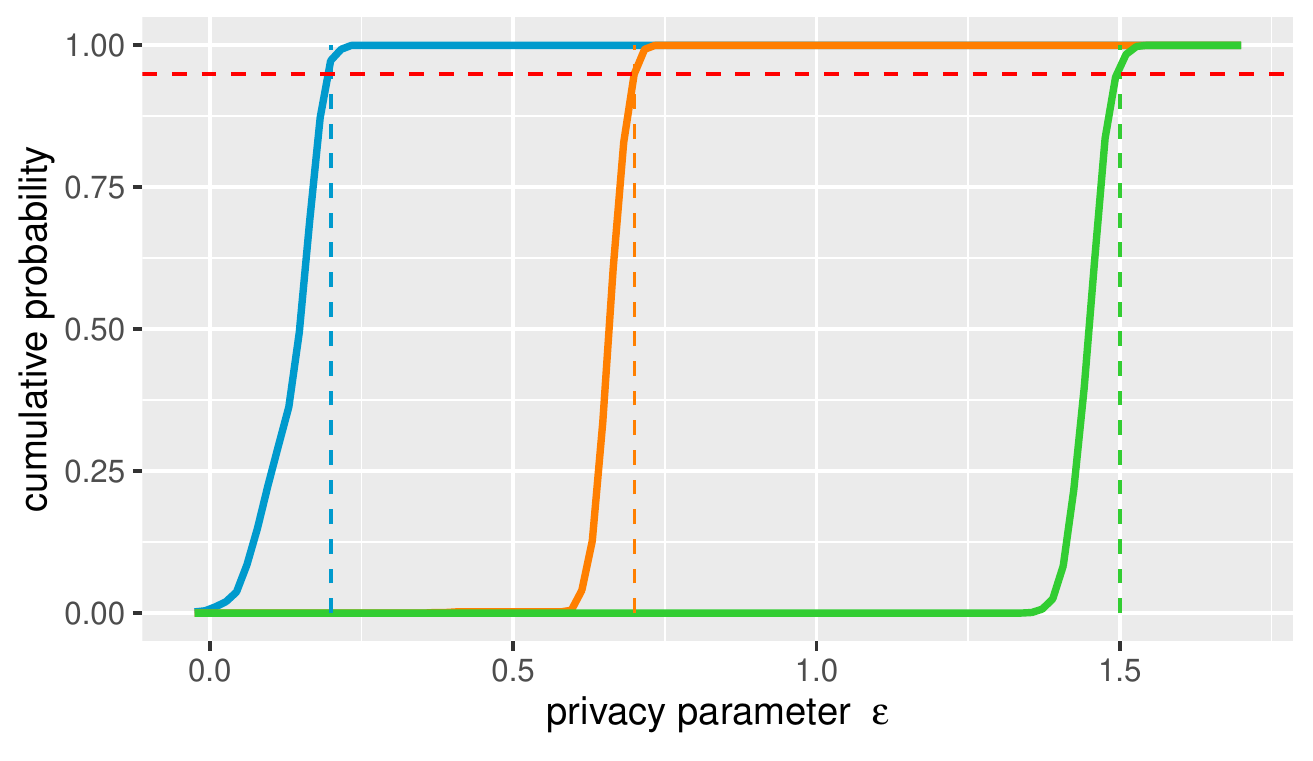}
\label{LB_exp_plot}
\end{subfigure}

\begin{subfigure}[c]{.49\linewidth}
\centering
\caption{\textbf{Sparse Vector Technique 5}}
\includegraphics[width=0.9\linewidth,height=140pt]{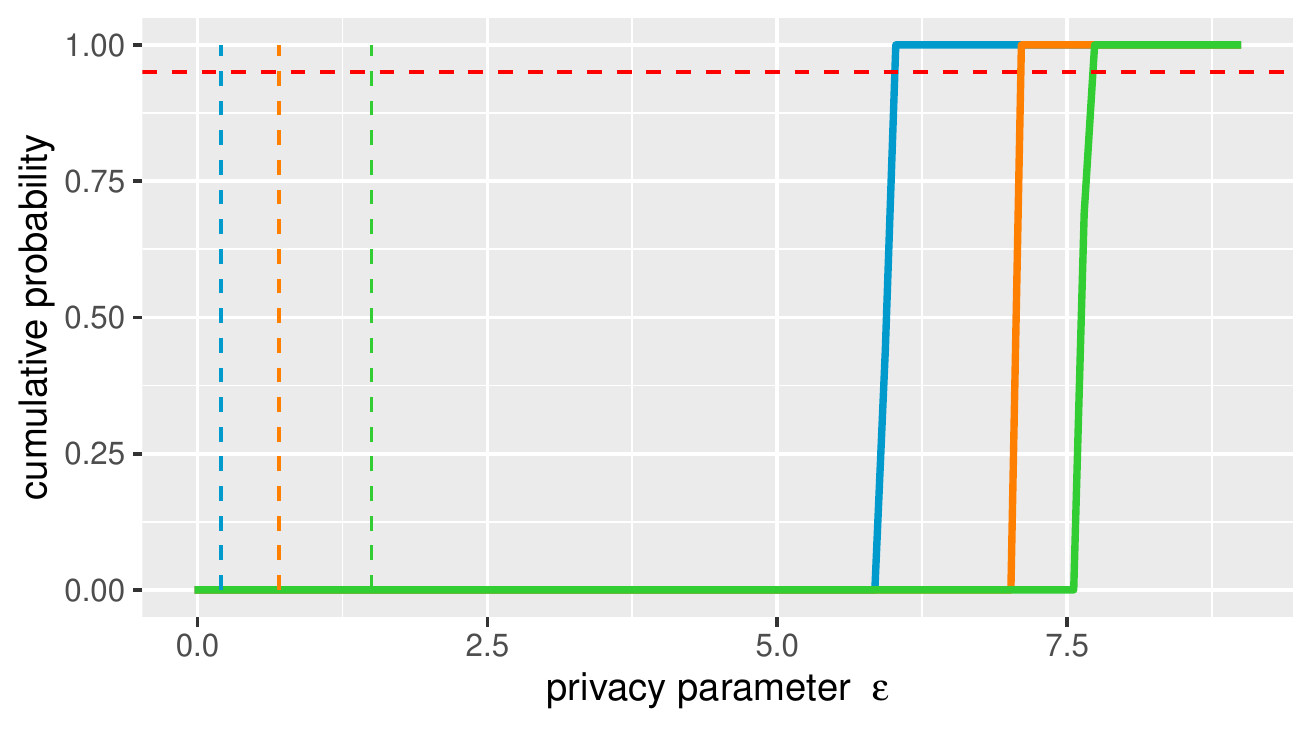}
\label{LB_max_plot}
\end{subfigure}
\quad
\begin{subfigure}[c]{.49\linewidth}
\centering
\caption{\textbf{Sparse Vector Technique 6}}
\includegraphics[width=0.9\linewidth,height=140pt]{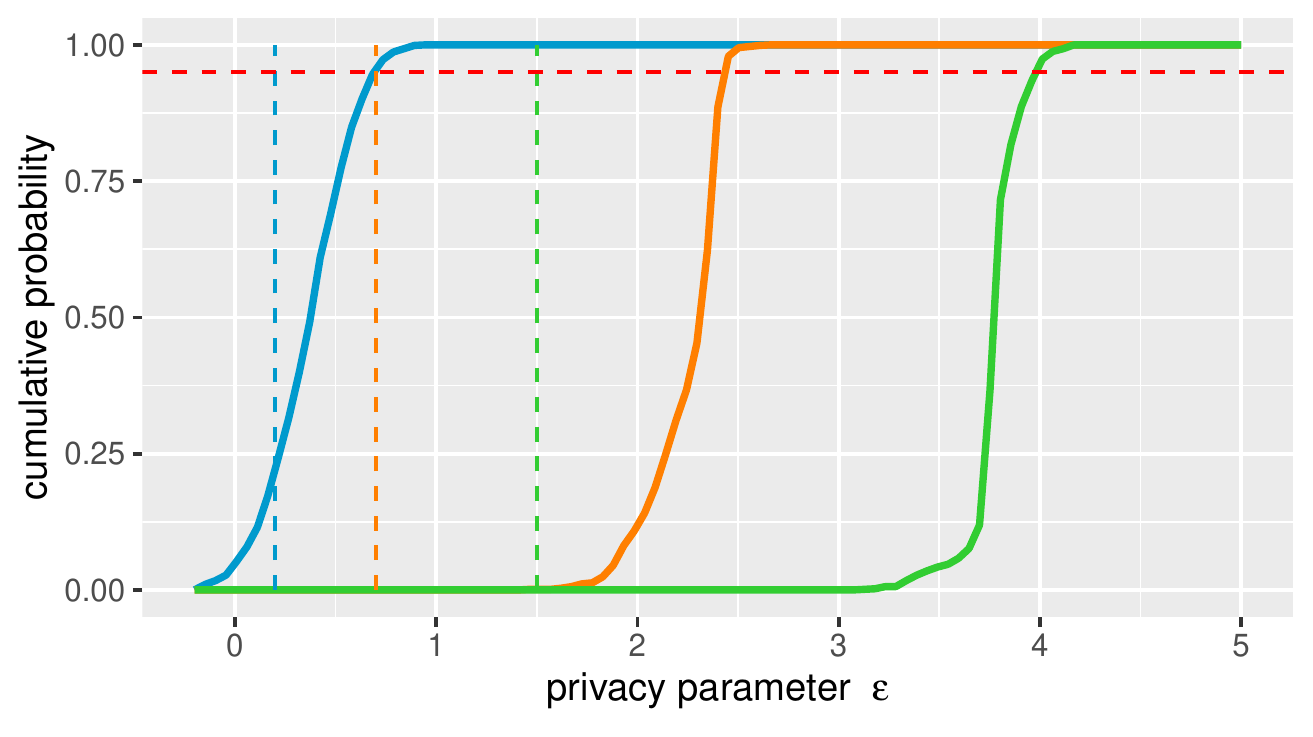}
\label{LB_exp_plot}
\end{subfigure}
\caption{Empirical distribution functions of the lower bound $LB$ in the high (blue), middle (orange) and low (green) privacy regime, generated by the MPL algorithm. The vertical lines (with corresponding colors) depict the targeted privacy levels, and the red horizontal line the confidence level of $95\%$.  \label{Figure_panel}} 
\end{figure*}

The \textbf{continuous Noisy Max} algorithm (see Algorithm 7,  \cite{StatDP}) has been discussed in Example \ref{example_1}. Here we use it to publish the maximum entry of a statistic $s \in [0,1]^k$. We consider the case $k=3$ and input statistics $s_b = (0,0,0)$ and $s_b'=(b/10,b/10,b/10)$ for $b=1,...,10$. Furthermore, we choose $C = [-1,1]$. 

The \textbf{Exponential Mechanism} provides a general principle for the construction of private algorithms. We consider a version where we privatize real numbers from the interval $[1,2]$, with non-negative outputs. More precisely, for a number $s \in [1,2]$ the output is sampled according to a continuous density proportional to $\exp(-\lambda |s-t|)$ for $t\ge 0$. Here $\lambda>0$ is a parameter determining the privacy level. Recall that this setup fits our (relaxed) notion of continuous algorithms discussed in Section \ref{Sec_3} (continuous density on the half-line). It is well known that using this construction, the exponential mechanism affords (at least) $2\lambda$-DP. We can however employ Theorem \ref{theorem_local_property} to derive the privacy parameter $\epsilon$ precisely:
$$
\epsilon = \lambda + \ln(2-\exp(-2\lambda))-\ln(2-\exp(-\lambda)).
$$
Notice that $\epsilon \approx 2\lambda$ for small $\lambda$. In the following simulations, we consider input statistics $s_b = 1$ and $s_b'=1+b/10$ for $b=1,...,10$ and choose $C = [0,2]$.

\begin{figure*}
\begin{subfigure}[c]{.49\linewidth}
\centering
\caption*{\textbf{Report Noisy Max}}
\includegraphics[width=0.9\linewidth,height=160pt]{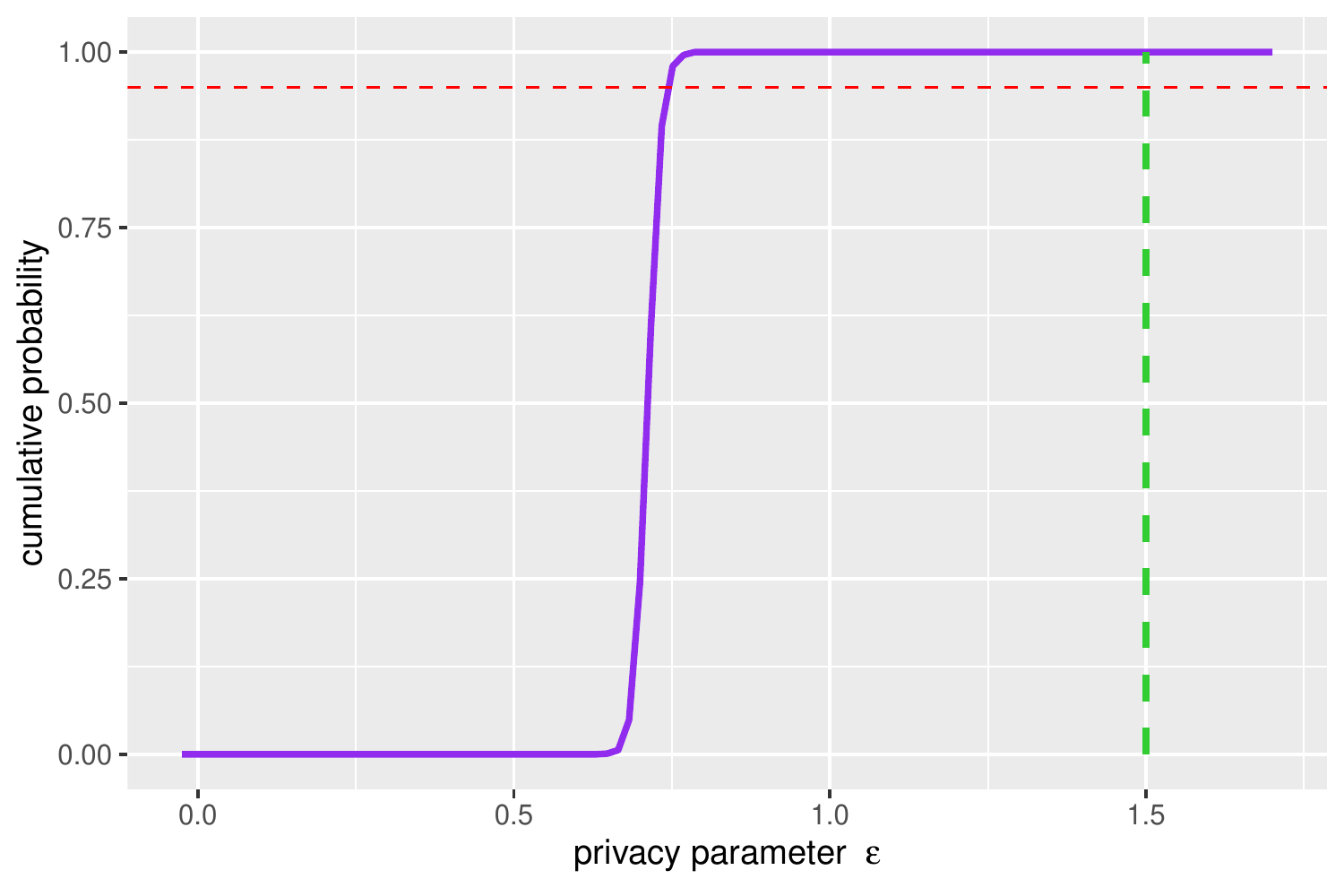}
\label{noisy_local_plot}
\end{subfigure}
\quad
\begin{subfigure}[c]{.49\linewidth}
\centering
\caption*{\textbf{Continuous Noisy Max}}
\includegraphics[width=0.9\linewidth,height=160pt]{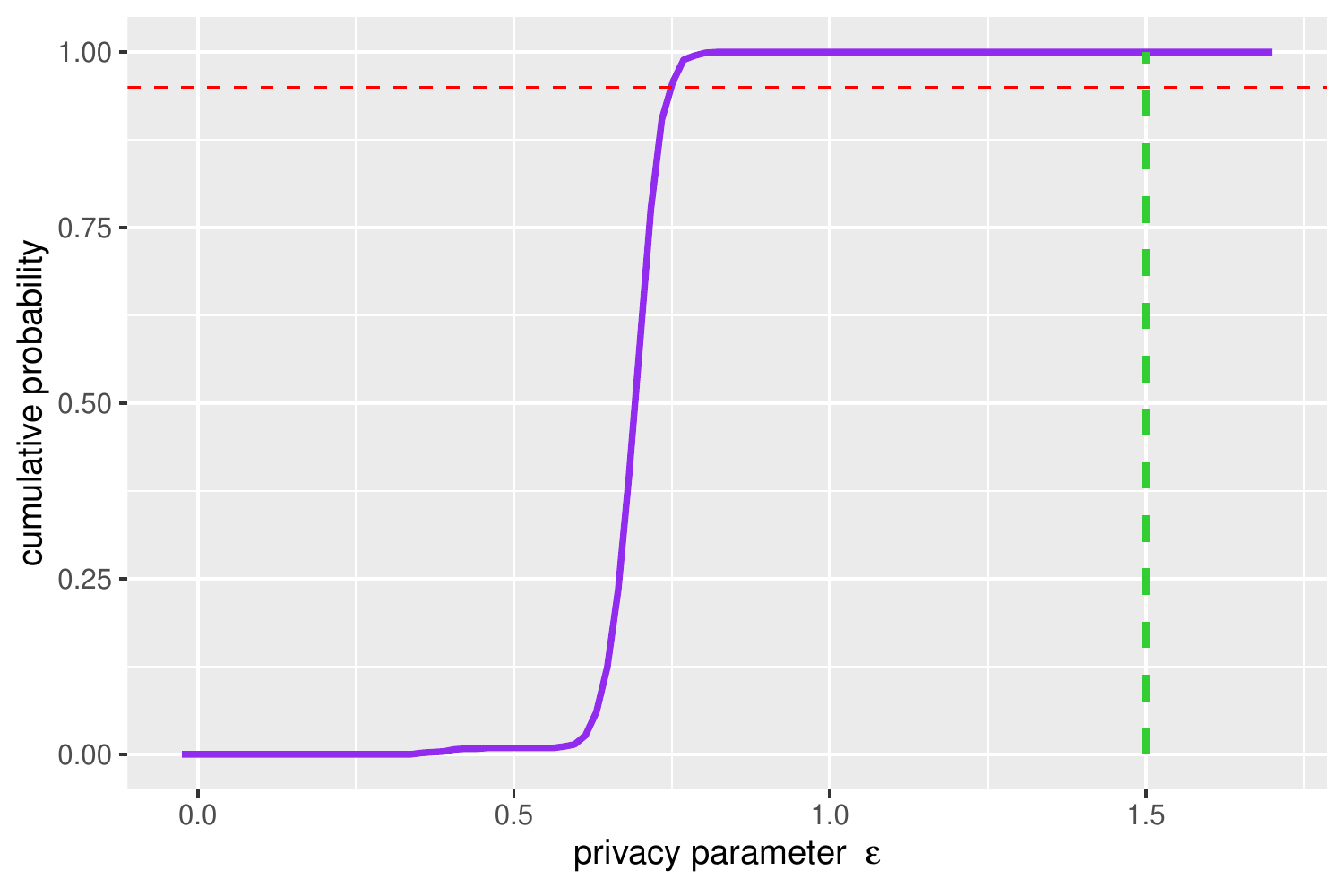}
\label{max_local_plot}
\end{subfigure}
\caption{Empirical distribution function of $LB$ for fixed databases.}  \label{Figure_eps_x}
\end{figure*}

\subsubsection*{\textbf{Experiment settings}} 
To study privacy violations, we employ the MPL algorithm described in Section \ref{subsec_43}. The sample sizes and floor in MPL are chosen as $n=2 \times 10^4$, $N=5 \times 10^4$ and $\tau = 10^{-3}$ for algorithms (a)-(d) (labels as in Figure \ref{Figure_panel}), i.e. all algortithms apart from the SVTs. For the SVTs we use larger sample sizes and a smaller floor with $n=10^5$, $N=5 \times 10^5$ and $\tau = 10^{-4}$. This choice of parameters is necessary as SVTs allow for extreme events (with low probability) that otherwise cause instabilities.

For the continuous algorithms, the kernel in KDE is the Gaussian Kernel (described in Appendix B) and the bandwidths in the first step of MPL are chosen by a pre-implemented selection rule in the “kdensity” package (both are the default options).

We examine each algorithm for different targeted privacy parameters $\epsilon_0 \in \{ 0.2, 0.7, 1.5 \} $, capturing the high, middle and low privacy regime respectively \cite{StatDP} (we adjust the targeted privacy level, e.g. by tuning the Laplace noise or changing $\lambda$ in the Exponential Mechanism).
Correctly designed algorithms meet their targeted privacy levels, i.e. $\epsilon = \epsilon_0$. Algorithms (a) - (f) fall into this category, with labels again as in Figure \ref{Figure_panel}. Notice that (f) is sometimes deemed “incorrect” in the literature \cite{StatDP}, as in its original design $\epsilon$ is only equal to the targeted level $\epsilon_0$ up to a constant (this simple scaling error has been corrected in our version). Algorithms (g) and (h) constitute incorrect algorithms that do not satisfy DP at all, i.e. $\epsilon = \infty$ \cite{Lyu2017}. Recalling \eqref{Eq_max_violation}, this especially points to privacy violations $\epsilon_{x,x'}$ that exceed the targeted privacy parameter $\epsilon_0$.

\subsubsection*{\textbf{Results}} 

In order to evaluate MPL, we consider the cumulative distribution function (cdf) of the lower bound $LB$ defined in \eqref{Eq_def_LB}. Recall that the cdf is defined for some $z \in \mathbb{R}$ as $\mathbb{P}(LB \le z)$. In Figure \ref{Figure_panel} we display a panel where each plot corresponds to one algorithm under investigation and each curve to the empirical cdf for a different choice of $\epsilon$ (each based on $1000$ simulation runs). This presentation is related to, but more informative than, a standard histogram and for details on the empirical cdf we refer to \cite{vandervaart1998}. 
It is also particularly transparent, as we report the results of $1000$ simulated lower bounds (instead of just a single one), giving insight into the variance of $LB$.
The dashed vertical lines (in the same color as the corresponding cdfs) indicate the targeted privacy parameters $\epsilon_0$ and the horizontal, red line the prescribed confidence level $1-\alpha$, where we have chosen $\alpha = 0.05$.

For the correct algorithms (a) - (f) an important feature of the empirical cdfs is their location. Note that evaluated in the targeted privacy parameter $\epsilon_0 = \epsilon$, the cdf describes the confidence level $\mathbb{P}(LB \le \epsilon)$, which according to our theory should approximately equal $1-\alpha$ (see Theorem \ref{theorem_3}). Therefore, we would expect our empirical cdfs to pass through the intersection of the horizontal confidence level and the vertical targeted privacy level. In most scenarios we observe that the prescribed confidence level is indeed well approximated, while sometimes it is slightly too large (corresponding to small values of $LB$). 

This tendency is inherent in the empirical study of DP and should not surprise us: To approximate $\epsilon$, one has to first select the right data pair out of $B$ pairs and then empirically maximize the privacy loss. Poor performance in either step biases estimates away from $\epsilon$ towards smaller values - a trend that has been observed in other empirical studies (see e.g. \cite{StatDP}, where the $p$-values are in each instance much higher than the prescribed level). 

A second performance measure for our correct algorithms is the ascent of the cdf in a neighborhood of $\epsilon$: In most of our simulations (a)-(f) we observe a rapid increase close to $\epsilon$, suggesting that $LB$ is a tight and reliable bound for $\epsilon$. 
In the case of SVT2 and SVT4 the ascent is slightly slower in the high privacy regime $\epsilon_0 = 0.2$, which hints at higher variance in $LB$ caused by smaller values of the discrete densities.
As for the incorrect algorithms (g) and (h), the feature that provides the most conclusive information on the performance of MPL is the location of the empirical cdfs. To be more exact, a lower bound $LB$ to the right of the targeted privacy parameter exposes a false privacy claim (this corresponds to a right-shift of the empirical cdf). 
We observe that $LB$ is usually sampled to the right of its targeted privacy  parameter $\epsilon_0$ (with almost certainty for (g) and in the middle and low privacy regime for (h)), often with a large margin. In the high privacy regime for (h), we sometimes observe $LB\le \epsilon_0$, due to increased variance.  In conclusion, the experiments confirm the performance of MPL with respect to flawed algorithms.

\subsubsection*{\textbf{Sample sizes and runtime}} 

After considering the statistical results of our experiments, we want to briefly discuss computational aspects. Our MPL algorithm relies on standard statistical tools that are provided by many programming languages such as \texttt{R}. It is therefore convenient to implement for users.
Confirming this ease of applicability, we have run our simulations on a standard desktop computer (3.4 GHz Intel Core i5 CPU, 4 cores, 16 GB RAM). Under the above conditions runtimes range from 10 seconds for the smaller sample sizes (used for algorithms (a)-(d)) to less than one minute for the larger sample sizes (used for algorithms (e)-(h)). The precise runtimes are reported in Table \ref{table_1} and are shorter than those given in \cite{DP-Sniper}, where the above algorithms are also analyzed (except for the exponential mechanism). Importantly, \cite{DP-Sniper} also rely on  a much more powerful machine, with 128 cores at 1.2GHz and 500 GB RAM.

Our gains in terms of runtime are mainly achieved by cutting sampling efforts. For instance, consider $B=10$ pairs of neighboring databases as input for the MPL algorithm and its counterpart in \cite{DP-Sniper},  DD-Search. Then the total sampling effort associated with one run of MPL  amounts to $5\times 10^5$ for the smaller samples (algorithms (a)-(d)) and $3 \times 10^6$ for the larger ones (algorithms (e)-(h)). This corresponds to $\approx 0.05 \%$ and $\approx 0.32 \%$ of the sample sizes that would be used by the DD-Search algorithm in \cite{DP-Sniper}. This means that we rely only on a small fraction of the data used in \cite{DP-Sniper}. \\[-2ex] 

\begin{table}[h]
\centering
\begin{tabular}{ |p{2.2cm}|p{1.2cm}|p{2cm}|p{1.2cm}|  }
\hline
 \multicolumn{4}{|c|}{Runtime in seconds} \\
 \hline
 \hline
 Alg. & runtime & Alg. & runtime\\
 \hline
 Laplace (a)   & 10.9    &SVT 2 (e)&   23.8\\
 Noisy Max (b) &  $ \,$ 4.7  & SVT 4 (f)  & 26.6\\
 Noisy Max (c) &10.5 &  SVT 5 (g) &  25.9\\
 Exponential (d) &   11.3  &  SVT 6 (h) & 57.3\\
 \hline
\end{tabular}
\caption{Runtimes for one run of the MPL algorithm on (a)-(h). Times are averaged over 10 simulation runs.}    \label{table_1}
\end{table}

\subsubsection*{\textbf{The data-centric privacy level for fixed databases}} 

As pointed out in  Section \ref{Sec_4}, we can use the MPL algorithm to determine the data-centric privacy guarantee for select databases defined in \eqref{Eq_def_epsx}.  We demonstrate this on both versions (discrete and continuous) of the Noisy Max algorithm. 

Regarding the discrete case, suppose we have a database $x$ that, given $6$ counting queries, evaluates to $0$ for each query, that is $q = q(x) = (0,0,0,0,0,0)$. Recalling our discussion of the query model, we know that any database $x'$ in the neighborhood of $x$ evaluates to a binary vector $q' \in \{ 0,1 \}^6$. This means that the entire neighborhood of $x$ can be exhausted by the collection of all such query pairs $(q,q')$. We set the privacy parameter $\epsilon = 1.5$ and run the MPL algorithm for Report Noisy Max on that collection of query pairs 1000 times. In Figure \ref{Figure_eps_x} (left panel) we plot the empirical cdf of $LB$ (purple), which exhibits a sharp rise, long before the global privacy parameter $\epsilon$ (vertical green line).
In view of our earlier results and given the exhaustive search of query pairs, we can be confident that the empirical cdf captures the data-centric privacy leakage $\epsilon_x$.
The plot suggests that the data-centric privacy parameter is only about half the size of $\epsilon$, confirming that the amount of privacy afforded to this specific database outstrips the worst case guarantee.

For the continuous case, we consider a database $x$ that produces the statistic $s = S(x) =(1/2, 1/2, 1/2)$ and assume that $S$ maps neighboring databases $x'$ anywhere on the unit cube $[0,1]^3$. Let $s' \in \{0, 1/2, 1\}^3$ (which forms an even grid of $27$ points on the unit cube). We can run MPL on the collection of statistics thus obtained. It can be shown by similar methods as employed in Example \ref{example_1}, that $\epsilon_{x,x'} = \epsilon_x$  is attained for databases $x'$ with $S(x')=s'=(0,0,0)$ or  $S(x')=s'=(1,1,1)$, both of which are covered by our grid. As for the discrete case, we observe that $\epsilon_x$ is about half the size of $\epsilon$ (see Figure \ref{Figure_eps_x}, right panel). In conclusion, the amount of privacy ceded to our specific databases $x$ in both examples is about twice as high as the global privacy parameter suggests (i.e. $\epsilon_x \approx \epsilon/2$).

\subsubsection*{\textbf{Estimation of data-specific privacy violations}} 

Up to this point we have focused on the lower bound $LB$, produced by the MPL algorithm. We now want to consider the estimation of data-specific privacy violations defined in \eqref{lower_bound}, which is the key novelty of our local approach and, as an integral part of MPL, has an outsize effect on the quality of $LB$.
We especially focus on the two continuous algorithms (Noisy Max and the Exponential Mechanism), where our estimator $\hat{\epsilon}_{x,x'}$ differs most noticeably from prior approaches by virtue of kernel density estimation.

\begin{figure}
\begin{subfigure}[c]{.48\linewidth}
\centering
\caption*{\textbf{Continuous Noisy Max}}
\includegraphics[width=1\linewidth,height=180pt]{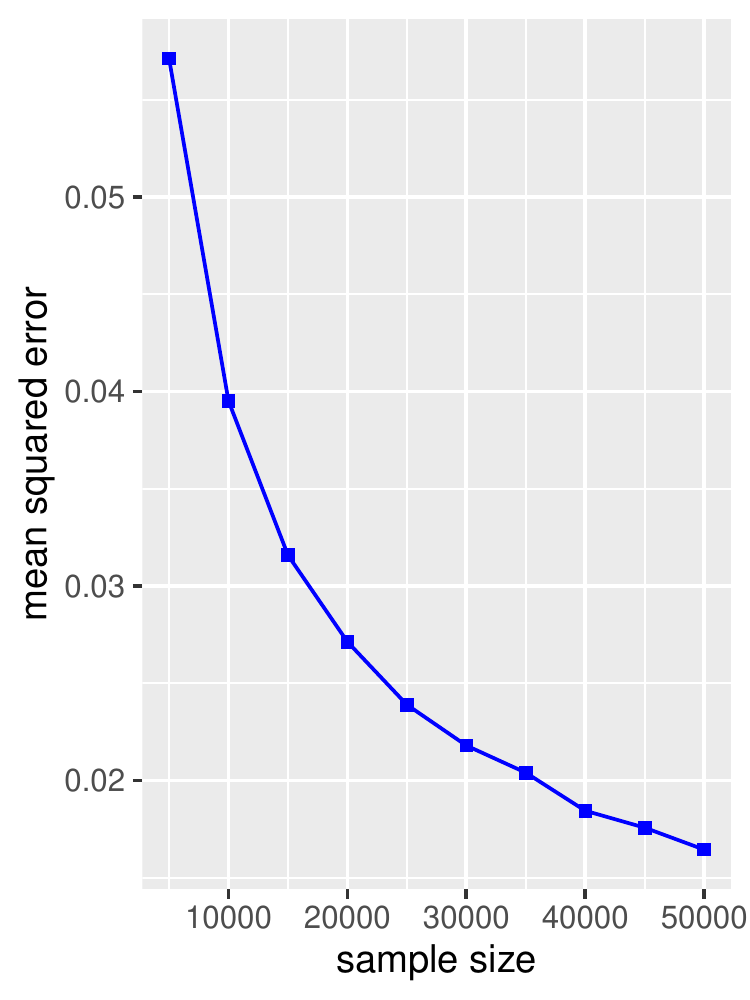}
\label{LB_noisy_plot}
\end{subfigure}
\quad
\begin{subfigure}[c]{.48\linewidth}
\centering
\caption*{\textbf{Exponential Mechanism}}
\includegraphics[width=1\linewidth,height=180pt]{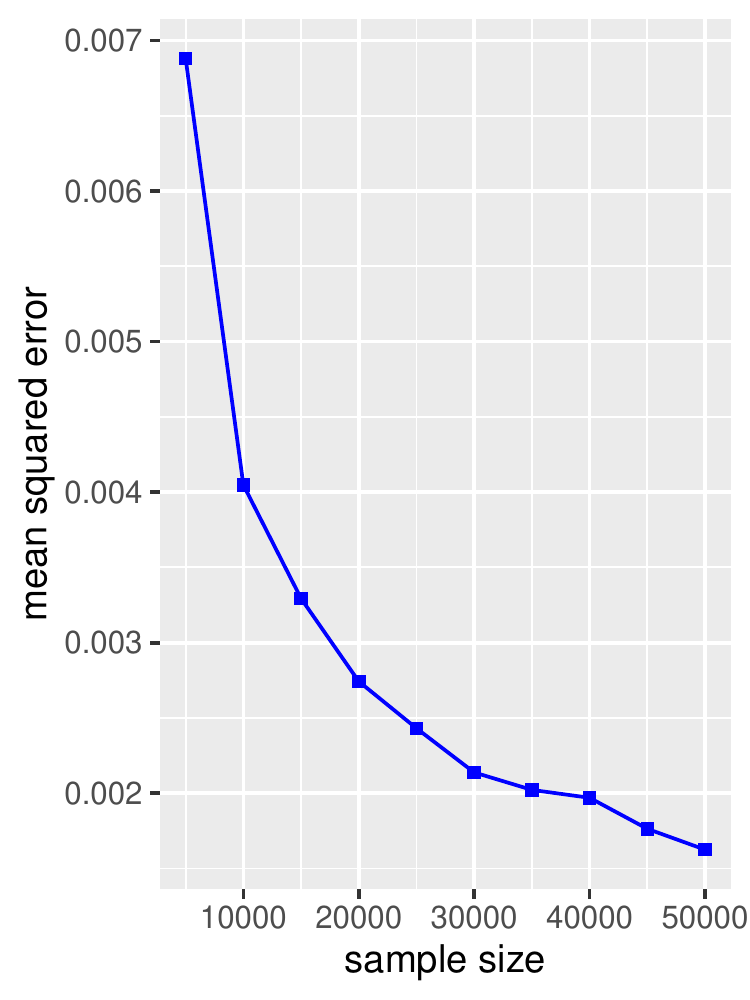}
\label{LB_svt_plot}
\end{subfigure}
\caption{Mean squared error $\mathbb{E}(\hat \epsilon_{x,x'}- \epsilon_{x,x'})^2 $ for different sample sizes $n$ and $\epsilon_{x,x'}=1.5$.}  \label{Figure_approximation}
\end{figure}

Regarding the Noisy Max algorithm, suppose we choose databases $x$ and $x'$ that produce statistics  $s = S(x) = (0,0,0)$ and $s' = S(x') =(1,1,1)$, and similarly for the Exponential Mechanism databases $x$ and $x'$ that result in $s=1$ and $s' = 2$. In both situations,  the choice of these databases  provokes a privacy violation $\epsilon_{x,x'}=\epsilon$ that is equal to the global privacy parameter, which we fix at $1.5$. 

To study the quality of the estimator $\hat{\epsilon}_{x,x'}$ based on $n$ observations, we consider the mean squared error $\mathbb{E}(\hat \epsilon_{x,x'}- \epsilon_{x,x'})^2 $ (approximated by $1000$ simulation runs) for both algorithms. In Figure \ref{Figure_approximation} we display the simulated errors for the two algorithms and different sizes of $n$. In both cases we observe for a sample size as moderate as $5000$ only small estimation errors (less than $4\%$ of the true $\epsilon$ for Noisy Max and less than $0.5 \%$ for the Exponential Mechanism) and the errors are less than half of this for $n=20000$ (which is used in our previous experiments). This shows that the strong performance of MPL can also be attributed to the precision of our local estimators for the data-specific privacy violations.

\section{Conclusion}

In this work, we have discussed a way to assess privacy with statistical guarantees in a black box scenario. In contrast to prior works, our approach relies on a local conception of DP that facilitates the estimation and interpretation of privacy violations by circumventing the problem of event selection. Besides quantification of the global privacy parameter, our methods can be used for a more refined analysis, measuring the amount of privacy ceded to a specific database. 
The findings of this analysis might not only help to understand existing algorithms better, but also aid the design of new privacy preserving mechanisms. This can, for instance, be algorithms that are tailored to provide  greater privacy to databases that require more protection.

\section*{Acknowledgments}
This work was supported by the Deutsche Forschungsgemeinschaft (DFG, German Research Foundation) under Germany's Excellence Strategy - EXC 2092 CASA - 390781972. We would also like to thank the anonymous reviewers for their fruitful comments and suggestions to improve this work.

\normalem

\bibliographystyle{IEEEtran}
\begin{small}
\bibliography{references}
\end{small}

\appendices

\section{Proofs and technical details} \label{App_1}

The appendix is dedicated to the mathematical details of our analysis: the definition of stochastic convergence, additional facts on the kernel $K$ in KDE, as well as the proofs of Proposition \ref{proposition_1} and Theorem \ref{theorem_3}.

\subsection{Stochastic Landau symbols and convergence in probability} \label{Appendix_convergence}

Let $(Z_n)_{n \in \mathbb{N}}$ be a sequence of random variables and $(a_n)_{n \in \mathbb{N}}$ a sequence of positive, real numbers. We now say that $Z_n = \mathcal{O}_P(a_n)$, if for every $\varepsilon>0$ there exists a (sufficiently large) $C>0$ s.t.
$$
  \limsup_{n \to \infty} \mathbb{P}(|Z_n|/a_n \ge C)<\varepsilon.
$$
Notice that analogous rules hold for the stochastic as for the deterministic Landau notation, such as $\mathcal{O}_P(a_n) = a_n\mathcal{O}_P(1)$ or, for another positive sequence  $(b_n)_{n \in \mathbb{N}}$, that $\mathcal{O}_P(a_n)+\mathcal{O}_P(b_n) = \mathcal{O}_P(a_n+b_n)$. 
Next we say that $Z_n = o_P(a_n)$, if for every (arbitrarily small) $c>0$ 
$$
  \lim_{n \to \infty} \mathbb{P}(|Z_n|/a_n \ge c)=0.
$$
Finally we say that for a constant $a \in \mathbb{R}$ it holds that $Z_n \to_P a$ if $|Z_n-a|=o_P(1)$. We say that $Z_n \to_P \infty$, if for any $C>0$ 
$$
  \lim_{n \to \infty} \mathbb{P}(Z_n \ge C)=1.
$$
For an extensive explanation of Landau symbols and convergence see \cite{bishop}.

\subsection{Kernel density estimation} \label{Appendix_KDE}

Recall the definition of a kernel $K$ as a continuous function $K: \mathbb{R}^d \to \mathbb{R}_{\ge 0}$ with $\int_{\mathbb{R}^d} K(u) du =1$. In our discussion, we make the following two regularity assumptions, which are taken from \cite{jiang2017} (Assumptions 2 and 3):

\begin{itemize}
    \item[(K1)] $K$ satisfies \textit{spherical symmetry}, i.e. there exists a non-increasing function $k: \mathbb{R}_{\ge 0} \to \mathbb{R}_{\ge 0}$, s.t.
    $
    K(u) = k(|u|)$ $\forall u \in \mathbb{R}^d.
    $
    \item[(K2)] $k$ has \textit{exponentially decaying tails}, i.e. there exist $\rho, C_\rho, t_0$, s.t. 
    $
    k(t) \le C_\rho \exp (-t^\rho), \,\, \forall t >t_0.
    $
\end{itemize}

A typical example of a kernel satisfying (K1) and (K2) is the \textit{Gaussian kernel}, which corresponds to the density function of a standard normal and is given for $d=1$ as 
$
    K(t) = \exp( -\frac{t^2}{2})/\sqrt{2 \pi}.
$
We use this kernel in our experiments to study continuous algorithms.

\subsection{Proof of Proposition \ref{proposition_1}}

We only show the proposition for the case of a continuous algorithm $A$ and only for $d=1$ (the case $d>1$ is a straightforward generalization).  The discrete case works by similar, but simpler techniques. Here, the central limit theorem can be employed to establish a uniform convergence rate of $\mathcal{O}_P(n^{-1/2})$ for the relative frequency estimator. 
By exploiting the differentiability of the logarithm, this rate of convergence can then be transferred to $\hat{\epsilon}_{x,x'}$.
The second identity in the discrete case follows as $\ell_{x,x'}(\hat t) = \epsilon_{x,x',C}$ with probability converging to one (which is not true in the continuous case). In the following, we restrict ourselves to the case where $\epsilon_{x,x',C} \in (0, \infty)$. Proving consistency in the remaining cases $\epsilon_{x,x',C} \in \{0, \infty\}$ is easier and therefore omitted.\\
We begin by defining two sets, that will be used extensively in our subsequent discussion: the argmax of the loss function
$$
\mathcal{M}:= \argmax_{t \in C} \ell_{x,x'}(t)
$$
and the closed $\zeta$-environment of $\mathcal{M}$
$$
U_\zeta (\mathcal{M}) := \{t \in C: \min_{t' \in \mathcal{M}} |t-t'|\le \zeta   \}.
$$
Notice that $\mathcal{M}$ is non-empty and closed. To see this, consider a sequence $(t_n)_{n \in \mathbb{N}} \subset C$, such that $\ell_{x,x'}(t_n) \to \sup_{t \in C}\ell_{x,x'}(t)$. Condition (C2) implies that there exists a limit point in $C$, where the maximum is attained. In particular $\mathcal{M} \neq \emptyset$. Similarly, we can show that $\mathcal{M}$ is closed: If $t$ is in the closure of $\mathcal{M}$, we can construct a sequence $(t_n)_{n \in \mathbb{N}}\subset \mathcal{M}$ with $t_n \to t$ and by Condition (C2) it follows that $t \in \mathcal{M}$. 

We now formulate an auxiliary result, that is the main stepping stone in the proof of Proposition \ref{proposition_1}.

\begin{lemma} \label{lemma_1}
Suppose that the assumptions of Proposition \ref{proposition_1} hold and $\epsilon_{x,x',C} \in (0, \infty)$. Then the following statements hold:
\begin{itemize}
    \item[i)] For any sufficiently small $\zeta>0$
    $$\sup_{t \in U_\zeta (\mathcal{M})}|\hat \ell_{x,x'}(t)- \ell_{x,x'}(t)|=\mathcal{O}_P \Big( \sqrt{\ln(n)} n^{-\frac{\beta}{2\beta+1}}\Big). $$
    \item[ii)] There exists a $\kappa=\kappa(\zeta)>0$ s.t.
    \begin{equation*} 
    \lim_{n \to \infty} \mathbb{P}\Big( \sup_{t \not \in  U_\zeta(\mathcal{M})} \hat \ell_{x,x'}(t) > \sup_{t \in C}\ell_{x,x'}(t) -\kappa \Big) = 0.
\end{equation*}
\end{itemize}
\end{lemma}

Let us verify that the Lemma indeed entails Proposition \ref{proposition_1}. We first show that for a small enough $\zeta>0$ it holds that
\begin{equation} \label{Eq_hat_t_in_U}
\lim_{n \to \infty} \mathbb{P}\Big( \hat t \in U_{\zeta}(\mathcal{M})\Big) =1.
\end{equation}
To see this we notice that according to Lemma \ref{lemma_1}, part ii) there exists a $\kappa>0$, s.t.
$$
 \sup_{t \not \in  U_\zeta(\mathcal{M})} \hat \ell_{x,x'}(t) \le  \sup_{t \in \mathcal{M}}\ell_{x,x'}(t) -\kappa +o_P(1).
$$
Here we have used $  \sup_{t \in \mathcal{M}}\ell_{x,x'}(t)=  \sup_{t \in C}\ell_{x,x'}(t)$. Combining this with part i) of the lemma we have
$$
 \sup_{t \not \in  U_\zeta(\mathcal{M})} \hat \ell_{x,x'}(t) \le  \sup_{t \in \mathcal{M}} \hat \ell_{x,x'}(t) -\kappa +o_P(1).
$$
As a consequence it holds with probability converging to $1$, that $\hat \ell_{x,x'}$ does not attain its maximum in $C \setminus U_\zeta (\mathcal{M})$ and conversely that \eqref{Eq_hat_t_in_U} holds. We now have for any $t^* \in \mathcal{M}$
\begin{align} \label{Eq_max_rates}
     |\hat \ell_{x,x'}(t^*)- \ell_{x,x'}(t^*)|= &\mathcal{O}_P \Big( \sqrt{\ln(n)} n^{-\frac{\beta}{2\beta+1}}\Big)\\
      |\hat \ell_{x,x'}(\hat t)- \ell_{x,x'}(\hat t)|  = & \mathcal{O}_P \Big( \sqrt{\ln(n)} n^{-\frac{\beta}{2\beta+1}}\Big), \nonumber
\end{align}
where we have used part i) of the Lemma and for the second rate additionally \eqref{Eq_hat_t_in_U}. Now, the first identity in Proposition \ref{proposition_1} (in the continuous case) follows by comparing the empirical and the true loss function at their respective argmaxes. For instance, supposing that $\hat \ell_{x,x'}(\hat t) \ge \ell_{x,x'}(t^*)$ holds, we have
\begin{align*}
  &  |\hat \epsilon_{x,x'} - \epsilon_{x,x',C} | = \ell_{x,x'}(\hat t) - \ell_{x,x'}(t^*) \\
  = & \ell_{x,x'}(\hat t) -\ell_{x,x'}(\hat t) + \ell_{x,x'}(\hat t) - \ell_{x,x'}(t^*) \\
  = & \mathcal{O}_P \Big( \sqrt{\ln(n)} n^{-\frac{\beta}{2\beta+1}}\Big) + [\ell_{x,x'}(\hat t) - \ell_{x,x'}(t^*)] \ge 0.
\end{align*}
Non-negativity follows from  $\hat \ell_{x,x'}(\hat t) \ge \ell_{x,x'}(t^*)$, while the decay rate in the second equality follows from \eqref{Eq_max_rates}. Since $[\ell_{x,x'}(\hat t) - \ell_{x,x'}(t^*)] $ is non-positive, it must also hold that 
$$
|\ell_{x,x'}(\hat t) - \ell_{x,x'}(t^*)| = \mathcal{O}_P \Big( \sqrt{\ln(n)} n^{-\frac{\beta}{2\beta+1}}\Big).
$$
Reversing the roles of empirical and true loss can be used to treat the case  $\hat \ell_{x,x'}(\hat t) \le \ell_{x,x'}(t^*)$.
Part ii) of the proposition also follows from \eqref{Eq_max_rates}, as
\begin{align*}
 &  |\epsilon_{x,x',C}-\ell_{x,x'}(\hat t)| = \ell_{x,x'}( t^*)-\ell_{x,x'}(\hat t)\\
  = &  [\ell_{x,x'}( t^*)-\hat \ell_{x,x'}(\hat t)]+[\hat\ell_{x,x'}(\hat t) -\ell_{x,x'}(\hat t)].
\end{align*}
In the first step we have used that $\epsilon_{x,x',C} =\ell_{x,x'}( t^*) \ge \ell_{x,x'}(\hat t)$ because $t^* \in \mathcal{M}$. We can now treat the two terms on the right separately. The first term in the square brackets decays at the desired rate according to Proposition \ref{proposition_1} part i) and the second part according to the second identity in \eqref{Eq_max_rates}. This shows Proposition \ref{proposition_1} in the continuous case. \\

We now show that Lemma \ref{lemma_1} holds. We begin with two technical observations: For any, sufficiently small $\zeta>0$ there exist positive constants $\kappa, \rho>0$, such that simultaneously
\begin{align}
    & \min_{t \in U_\zeta (\mathcal{M})} f_x(t) \land f_{x'}(t) \ge \rho >0 \label{eq_c1}\\
    &  \sup_{t \in C \setminus U_\zeta (\mathcal{M})} \ell_{x,x'}(t) < \sup_{t \in C} \ell_{x,x'}(t) -\kappa \label{eq_c2},
\end{align}
where “$a \land b$” denotes the minimum of two numbers $a$ and $b$.\\
We begin by proving \eqref{eq_c1}: For all $t \in \mathcal{M}$ it holds that $f_x(t) \land f_{x'}(t)>0$ (otherwise the assumption $\sup_{t \in C} \ell_{x,x'}(t) \in (0, \infty)$ would be violated). Now $f_x \land f_{x'}$ is a continuous function on the closed (thus compact) set $\mathcal{M}$ and it therefore attains its (positive) minimum. Therefore, for some $\tilde \rho>0$ it holds that
$
\min_{t \in \mathcal{M}} f_{x}(t) \land f_{x'}(t) \ge \tilde \rho.
$
Now let $t \in U_\zeta(\mathcal{M})$ and $\tilde t \in \mathcal{M}$, s.t. $|t- \tilde t| \le \zeta$. According to
(C1) it holds that 
\begin{align*}
  & f_{x}(t) \land f_{x'}(t) 
   \\ 
   \ge  & f_{x}(\tilde t) \land f_{x'}(\tilde t) - |f_{x}( t) \land f_{x'}( t)  -f_{x}(\tilde t) \land f_{x'}(\tilde t) | \\
  \ge  & \tilde \rho - a |\tilde t - t|^\beta \ge \tilde \rho - a \zeta^\beta.
\end{align*}
Here we have used for the second inequality that the minimum of two $\beta$-Hölder continuous functions is again $\beta$-Hölder (where we have called the constant $a$). In the last step we have used that $|t- \tilde t| \le \zeta$. It is now obvious that with sufficiently small $\zeta$, say $\zeta< (\tilde \rho/(2a))^{1/\beta}$, it follows \eqref{eq_c1} with $\rho:=\tilde\rho/2$.\\
Next we show \eqref{eq_c2}. Suppose \eqref{eq_c2} was wrong. Then there must exist a sequence $(t_n)_{n \in \mathbb{N}} \subset C \setminus U_\zeta(\mathcal{M})$ s.t. $\ell_{x,x'}(t_n) \to \sup_{t \in C}\ell_{x,x'}(t)$. According to (C2) there exists a limit point $t^*$, where the maximum is attained. By definition $t^* \in \mathcal{M}$. This however is a contradiction to the fact, that $|t_n - t^*|>\zeta$ for all $n \in \mathbb{N}$, showing \eqref{eq_c2}. In the following we assume that $\kappa, \rho, \zeta$ are chosen such that \eqref{eq_c1} and \eqref{eq_c2} hold.

We now prove part i) of Lemma \ref{lemma_1}. To show this, we first notice that for any fixed $\rho' \in (0, \rho)$ it holds that 
\begin{equation} \label{Eq_lowerbound_KDE}
    \lim_{n \to \infty} \mathbb{P}\Big( \tilde f_{x}(t) \land \tilde f_{x'}(t) > \rho': \forall t \in U_\zeta (\mathcal{M})\Big) = 1,
\end{equation}
where $\tilde f_{x}(t), \tilde f_{x'}(t) $ are the KDEs defined in \eqref{Eq_def_KDE}, Section \ref{Subsec_2_B}. \eqref{Eq_lowerbound_KDE} is a direct consequence of the uniform consistency of KDEs (see \eqref{Eq_KDE_rate_2}). Now recall the definition of the truncated KDE $\hat f_{x} :=\tilde f_{x} \lor \tau$. Since $\tau \to 0$ and \eqref{Eq_lowerbound_KDE} holds, it follows for all $t \in U_\zeta (\mathcal{M})$ simultaneously that $\hat f_{x}(t) =\tilde f_{x}(t)$, with probability converging to $1$. Consequently, the definition of the empirical loss implies with probability converging to $1$
$$
    \hat \ell_{x,x'}(t) = |\ln(\tilde f_{x}(t))-\ln( \tilde f_{x'}(t)) |, \quad  \forall t \in U_\zeta (\mathcal{M}).
$$
This means that to establish part i) of the Lemma, it suffices to show
\begin{align*}
    & \big||\ln(\tilde f_{x}(t))-\ln( \tilde f_{x'}(t)) |\\
    &-|\ln( f_{x}(t))-\ln(  f_{x'}(t)) |\big| =\mathcal{O}_P \Big( \sqrt{\ln(n)} n^{-\frac{\beta}{2\beta+1}}\Big).
\end{align*}
By the triangle inequality we can show 
the desired rate separately for $|\ln(\tilde f_{x}(t))-\ln( f_{x}(t)) |$ and $|\ln(\tilde f_{x'}(t))-\ln( f_{x'}(t)) |$. We restrict ourselves to the first term (the second one follows by analogous arguments). By the mean value theorem it follows that
\begin{equation} \label{Eq_bound_mean_value}
    |\ln(\tilde f_{x}(t))-\ln( f_{x}(t))| = \frac{|\tilde f_{x}(t)-f_{x}(t)|}{\xi(t)},
\end{equation}
where $\xi(t)$ is a number between $\tilde f_{x}(t), f_{x}(t)$. The numerator is of order
\begin{equation} \label{Eq_numerator}
    \sup_{t} |\tilde f_{x}(t)-f_{x}(t)| = \mathcal{O}_P \Big( \sqrt{\ln(n)} n^{-\frac{\beta}{2\beta+1}}\Big),
\end{equation}
where we have used the uniform approximation of kernel density estimators, from \eqref{Eq_KDE_rate_2}. The denominator is bounded away from $0$, with probability converging to $1$, as the bound
\begin{equation}\label{Eq_denominator}
\xi(t) \ge f_{x}(t) - |\tilde f_{x}(t)-f_{x}(t)| \ge \rho - o_P(1),
\end{equation}
holds uniformly for $t \in U_\zeta(\mathcal{M})$. Here we have used the lower bound \eqref{eq_c1} of the density $f_{x}$ on $U_\zeta(\mathcal{M})$. Together \eqref{Eq_numerator} and \eqref{Eq_denominator} imply the desired rate for the right side of \eqref{Eq_bound_mean_value}. By our above arguments, this shows part i) of Lemma \ref{lemma_1}.

Next we prove part ii) of Lemma \ref{lemma_1}.
Let us therefore define pointwise in $t$ the truncated density
$$
f_x^{(\tau)} (t) := \begin{cases} f_x (t), \quad if \quad \hat f_x(t)>\tau,\\
\tau, \quad\quad\,\,\, else \end{cases}
$$
and analogously the function $f_{x'}^{(\tau)}$. Therewith define the truncated loss
\begin{align} \label{Eq_delta_ell}
\ell^{(\tau)}_{x,x'}(t) := |\ln(f_x^{(\tau)} (t))-\ln(f_{x'}^{(\tau)}(t))|.
\end{align}
By definition it holds for any $\tau>0$ and any $t$, that $\ell_{x,x'}(t)\ge \ell^{(\tau)}_{x,x'}(t)$ (“$=$” if $\hat f_x(t), \hat f_{x'}(t)>\tau$ and “$\ge$” else). Now for any $t \in C \setminus U_\zeta(\mathcal{M})$ we consider the following decomposition
\begin{equation} \label{Eq_A_decomposition}
     \sup_{s \in C}\ell_{x,x'}(s) - \hat \ell_{x,x'}(t) = A_1+A_2+A_3+A_4,
\end{equation}
\begin{align*}
    \textnormal{where} \quad A_1 := & \sup_{s \in C}\ell_{x,x'}(s) - \sup_{s \in C \setminus U_\zeta(\mathcal{M})}\ell_{x,x'}(s)\\
    A_2 := & \sup_{s \in C \setminus U_\zeta(\mathcal{M})}\ell_{x,x'}(s) - \sup_{s \in C \setminus U_\zeta(\mathcal{M})}\ell^{(\tau)}_{x,x'}(s)\\
    A_3 := &\sup_{s \in C \setminus U_\zeta(\mathcal{M})}\ell^{(\tau)}_{x,x'}(s) - \ell^{(\tau)}_{x,x'}(t)\\
    A_4 := & \ell^{(\tau)}_{x,x'}(t) - \hat \ell_{x,x'}(t).
\end{align*}
Now $A_1 \ge \kappa$ holds according to \eqref{eq_c2}. Furthermore $A_2 \ge 0$ due to the inequality $\ell_{x,x'}(s)\ge \ell^{(\tau)}_{x,x'}(s)$ and $A_3 \ge 0$ because $t \in C \setminus U_\zeta(\mathcal{M})$. Finally we turn to $A_4$ and show that it is uniformly in $t$ of order $o_P(1)$. Using the triangle inequality, we can upper bound $A_4$ by
\begin{align*}
    |\ln( f_{x}^{(\tau)}(t))-\ln( \hat f_{x}(t)) | + |\ln( f_{x'}^{(\tau)}(t))-\ln( \hat f_{x'}(t)) |.
\end{align*}
Both terms can be treated analogously and so we focus on the first one. If $ \hat f_{x}(t) \le \tau$ it is equal to $0$ and thus we consider the case where $ \hat f_{x}(t) > \tau$. According to the mean value theorem 
\begin{equation} \label{Eq_remainder_A4}
|\ln( f_{x}^{(\tau)}(t))-\ln( \hat f_{x}(t)) | = \frac{|f_{x}^{(\tau)}(t)-\hat f_{x}(t)|}{\xi'(t)},
\end{equation}
where $\xi'(t)$ lies between $f_{x}^{(\tau)}(t)$ and $\hat f_{x}(t)$. Just as before, the numerator is uniformly of order 
$$
\sup_{t \in C} |f_{x}(t)-\tilde f_{x}(t)| =\mathcal{O}_P \Big( \sqrt{\ln(n)} n^{-\frac{\beta}{2\beta+1}}\Big),
$$
and the denominator is (asymptotically) bounded away from $0$, as
$$
\xi'(t) = \hat f_{x}(t) +\mathcal{O}_P(\sup_{t \in C} |f_{x}(t)-\tilde f_{x}(t)|) \ge \tau +o_P(\tau).
$$
In both cases we have used that if $\hat f_x(t)>\tau$ we have $f_{x}^{\tau}(t)-\hat f_{x}(t) = f_{x}(t)-\tilde f_{x}(t)$.
Furthermore we have used for the denominator the approximation rate \eqref{Eq_KDE_rate_2} and that according to (C3)
$$
\mathcal{O}_P \Big( \sqrt{\ln(n)} n^{-\frac{\beta}{2\beta+1}}\Big) =  o_P(\tau).
$$
These arguments imply that the right side of \eqref{Eq_remainder_A4} is uniformly in $t$ of order $o_P(\tau)/[\tau +o_P(\tau)]= o_P(1)$. By our above arguments we now have
$
A_1+A_2+A_3+A_4 \ge \kappa +o_P(1),
$
which implies by \eqref{Eq_A_decomposition} part ii) of Lemma \ref{lemma_1} (if we replace $\kappa$ by $2 \kappa$ in the above calculations).

\subsection{Proof of Theorem \ref{theorem_3}}\label{Appendix_Sec_B}

As with Proposition \ref{proposition_1}, we only show Theorem \ref{theorem_3} for continuous algorithms and $d=1$ (extensions to $d>1$ are straightforward). The proof rests on the asymptotic normality of $\hat \ell_{x_{max},x_{max}'}^*(\hat t_{max})$, where the point $\hat t_{max}$ and the randomness in the estimator $\hat \ell_{x_{max},x_{max}'}^*$ are independent. In the discrete case, the proof is much simpler, as $\hat t_{max}$ is eventually an element of the argmax of $\ell_{x_{max},x_{max}'}$ and hence it is easy to establish an asymptotically vanishing bias. 
This is not so in the continuous case, where $\hat t_{max}$ is only close to the argmax (as we have seen above) and the bias has to be controlled by an undersmoothing procedure.\\
In the following proof, we confine ourselves to part i) of the theorem (as the convergence in part ii) follows by similar but simpler techniques). For clarity of presentation, we will assume that there exists a unique $b^* \in \{1,...,B\}$, s.t.
\begin{equation} \label{Eq_def_b*}
    \epsilon_{x_{b^*},x'_{b^*},C} = \max(\epsilon_{x_1,x'_1,C},...,\epsilon_{x_B,x'_B,C}).
\end{equation}
Recall that the MPL algorithm consists of two steps: First the algorithm creates $B$ pairs of samples with $n$ elements each, to approximate $\epsilon_{x_b,x_b',C}$ by $\hat \epsilon_{x_b,x_b'}$. According to Proposition \ref{proposition_1}, these estimates are consistent and therefore with probability converging to $1$ it holds that $b_{max}=b^*$ (where $b_{max}$ 
is an estimator defined in the MPL algorithm and $b^*$ is defined in \eqref{Eq_def_b*}). For simplicity  we will subsequently  assume that $(x_{max}, x_{max}')= (x_{b^*}, x_{b^*}')$ (formally we can do this by conditioning of the event $\{ b_{max}=b^* \}$).
Next recall that from the first step of MPL we get empirical estimates $\hat \ell_{x_{max}, x_{max}'}$ of the loss function and $\hat t_{max}$ of the location of maximum privacy violation. These estimates are based on samples $X_1,...,X_n \sim f_{x_{max}}$, $Y_1,...,Y_n \sim f_{x_{max}'}$. We will use these esimators in our subsequent discussion and it is important to keep them distinct from the randomness in the second part of the algorithm.\\
In the second step, MPL generates fresh samples of size $N$ $X_1^*,...,X_N^* \sim f_{x_{max}}$, $Y_1^*,...,Y_N^* \sim f_{x_{max}'}$. The corresponding density estimates, generated by the TKDE algorithm are denoted by $\hat f_{x_{max}}^*$ and $\hat f_{x_{max}'}^*$ (to distinguish them from the estimators from the first step of the algorithm). Notice that these density estimators use the same kernel $K$ as in the first step, but bandwidth $h_{max}$ of a smaller size (the asymptotic rate is described in Condition (C4)). Correspondingly we define the loss based on the $*$-samples 
$$
\hat \ell_{x_{max}, x_{max}'}^*(t) := |\hat f_{x_{max}}^*(t)-\hat f_{x_{max}'}^*(t)|.
$$
We point out that by the choices of $n,N$ and the bandwidth $h_{max}$  (see Condition  (C4)) it holds that 
\begin{equation}\label{Eq_van_bias_1}
 \sqrt{\ln(n)} n^{-\frac{\beta}{2\beta+1}} = o\Big( \frac{1}{\sqrt{Nh_{max}}}\Big).
\end{equation}
Now consider the decomposition
\begin{align} \label{Eq_B_decomposition}
   & \sqrt{N h_{max}} \big( \sup_{t \in C}\ell_{x_{max},x_{max}'}(t)-\hat \ell_{x_{max},x_{max}'}^*(\hat t_{max})\big)\\
   =:&B_1+B_2+B_3 \nonumber
\end{align}
where
\begin{align*}
    B_1 := &  \sqrt{N h_{max}} \big( \sup_{t \in C}\ell_{x_{max},x_{max}'}(t) - \hat \ell_{x_{max},x_{max}'}(\hat t_{max})\big) \\
    B_2 := &   \sqrt{N h_{max}} \big( \hat \ell_{x_{max},x_{max}'}(\hat t_{max})-  \ell_{x_{max},x_{max}'}(\hat t_{max})\big)\\
    B_3 := &   \sqrt{N h_{max}}  \big( \ell_{x_{max},x_{max}'}(\hat t_{max})-\hat \ell_{x_{max},x_{max}'}^*(\hat t_{max})\big).
\end{align*}
According to Proposition \ref{proposition_1} together with \eqref{Eq_van_bias_1} it follows that $B_1, B_2 =o_P(1)$. Thus to show weak convergence of \eqref{Eq_B_decomposition} (which is key to our asymptotic result) we can show weak convergence of $B_3$.\\
In order to study $B_3$ we consider the more general object
$$
G(t) := \sqrt{N h_{max}}  \big( \ell_{x_{max},x_{max}'}( t)- \hat \ell_{x_{max},x_{max}'}^*( t)\big)
$$
which is defined for any $t \in U_\zeta (\mathcal{M})$ (for some small enough, fixed $\zeta$ s.t. \eqref{eq_c1} and \eqref{eq_c2} hold), where from now on
$$
\mathcal{M}:= \argmax_{t \in C} \ell_{x_{max},x_{max}'}(t).
$$
 We now notice that with probability converging to $1$ it holds for all $t \in U_\zeta (\mathcal{M})$ that
\begin{align} \label{Eq_same_sign}
    & sign(\ln(\hat f_{x_{max}}^*(t))-\ln(\hat f_{x_{max}'}^*(t)))\\
    =& sign( \ln(f_{x_{max}}(t))- \ln(f_{x_{max}'}(t))). \nonumber
\end{align}
This follows because the density estimators are uniformly consistent (see Section \ref{Subsec_2_B}, equation \eqref{Eq_KDE_rate_1}), together with boundedness away from $0$ on $U_\zeta(\mathcal{M})$ (see \eqref{eq_c1}). \\
For simplicity of presentation, we subsequently assume that the signum on the right side of \eqref{Eq_same_sign} is always $1$. This means that with probability converging to $1$
\begin{align*}
    G(t) = & \sqrt{N h_{max}} \big( [\ln(\hat f_{x_{max}}^*(t))-\ln(f_{x_{max}}(t))]\\
    & \quad \quad \quad \quad -[\ln(\hat f_{x_{max}'}^*(t))- \ln(f_{x_{max}'}(t))]\big).
\end{align*}
By the mean value theorem we can transform the right side to
$$
\sqrt{N h_{max}} \Big( \frac{\hat f_{x_{max}}^*(t)- f_{x_{max}}(t))}{\xi_1(t)}-\frac{\hat f_{x_{max}'}^*(t)- f_{x_{max}'}(t)}{\xi_2(t)}\Big).
$$
Here $\xi_1(t)$ lies between $\hat  f_{x_{max}}^*(t)$ and $f_{x_{max}}(t)$, and $\xi_2(t)$ between  $\hat  f_{x_{max}'}^*(t)$ and $f_{x_{max}'}(t)$. We now focus on the fraction of densities in $x_{max}$ (the other one is analyzed step by step in the same fashion). Using \eqref{eq_c1} and the uniform consistency of the density estimates it is a simple calculation to show that 
$$
\frac{\hat f_{x_{max}}^*(t)- f_{x_{max}}(t))}{\xi_1(t)} =  \frac{\hat f_{x_{max}}^*(t)- f_{x_{max}}(t)}{f_{x_{max}}(t)}+Rem,
$$
where $Rem$ is a (negligible) remainder of size $o_P(1/\sqrt{N h_{max}})$ (here we have applied the same techniques as in the discussion of \eqref{Eq_bound_mean_value}). We can rewrite the fraction on the right side as follows
\begin{align*}
&\frac{\hat f_{x_{max}}^*(t)- f_{x_{max}}(t)}{f_{x_{max}}(t) }\\
=&\frac{1}{N f_{x_{max}}(t)} \sum_{i=1}^N  \Big[ h_{max}^{-1}K\Big(\frac{t-X_i^*}{h_{max}} \Big)- f_{x_{max}}(t) \Big].
\end{align*}
By standard arguments it is now possible to replace $f_{x_{max}}(t)$ in the sum by $\mathbb{E} h_{max}^{-1}K\big(\frac{t-X_i^*}{h_{max}}\big)$, while only incurring a (uniformly in $t$) negligible error. More precisely:
\begin{align*}
    & \mathbb{E}h_{max}^{-1} K \Big(\frac{t- X_i^* }{h_{max}}\Big) = \int h_{max}^{-1}K \Big(\frac{t-s}{h_{max}}\Big) f_{x_{max}}(s)ds \\
    =&  \int K(s) f_{x_{max}}(sh_{max}+t) ds \\
    = & f_{x_{max}}(t) +\int K(s) |f_{x_{max}}(sh_{max}+t)-f_{x_{max}}(t)| ds \\
    = & f_{x_{max}}(t)+ \mathcal{O}(|h_{max}|^\beta)
\end{align*}
Here we have used symmetry of the kernel (K1) in Appendix B) in the second and Hölder continuity of order $\beta$ in the last equality (see Assumption (C1); for a definition of Hölder continuity recall \eqref{Eq_Hoelder}). We also notice that $\mathcal{O}(|h_{max}|^\beta) = o_P(1/\sqrt{N h_{max}})$, which makes the remainder asymptotically negligible. By similar calculations we can show that
\begin{align} \label{Eq_variance}
& \mathbb{V}ar \Big( h_{max}^{-1}K\Big(\frac{t-X_i^*}{h_{max}}\Big) \Big) \\
=& h_{max}^{-1} f_{x_{max}}(t) \int K^2(y) dy + Rem_2, \nonumber
\end{align}
where $Rem_2$ is a remainder of negligible order. We can use the same considerations for $f_{x_{max}'}$ to rewrite 
\begin{align*}
    G(t) = \frac{1}{\sqrt{N}}\sum_{i=1}^N \{Z_i(t)-\mathbb{E}Z_i(t)\} + o_P(1),
\end{align*}
where 
$$
Z_i(t) = h_{max}^{-1/2} \Big[ K\Big( \frac{t-X_i^*}{h_{max}}\Big)+ K\Big( \frac{t-Y_i^*}{h_{max}}\Big) \Big].
$$
All variables $Z_i$ are i.i.d. and, according to \eqref{Eq_variance} (and analogous calculations for $f_{x_{max}'}$), asymptotically have variance
$$
\sigma^2(t) := \int K^2(y) dy \big([f_{x_{max}}(t)]^{-1} +  [f_{x_{max}'}(t)]^{-1}\big),
$$
 Now define the estimator 
\begin{align*}
\hat \sigma^2(t)& := \int K^2(y) dy  \big([\hat f_{x_{max}}^*(t)]^{-1} + [\hat f_{x_{max}'}^*(t)]^{-1}\big),
\end{align*}
which is identical to $\hat \sigma^2_{N}$ in MPL for $t=\hat t_{max}$.
By similar techniques as before, we can show that $\hat \sigma^2(t)$ is uniformly (for $t \in U_\zeta(\mathcal{
M})$) consistent for $\sigma^2(t)$. As a consequence, we have
$
G(t)/\hat \sigma(t) = S(t) + o_P(1),
$
where 
\begin{equation} \label{Eq_def_S}
    S(t) :=\frac{1}{\sqrt{N}}\sum_{i=1}^N \tilde Z_i(t)
\end{equation}
and $\tilde Z_i(t) := \{Z_i(t)-\mathbb{E}Z_i(t)\}/\sqrt{\mathbb{V}ar(Z_i)}$. We can now prove the identity \eqref{Eq_level}: 
First notice that 
\begin{align} \label{Eq_level_proof_1}
    & \mathbb{P}(LB \le \epsilon_{C}^*) = \mathbb{P}(LB \le \epsilon_{x_{max},x_{max}',C})\\
    =&\mathbb{P}\Big(    \hat \ell^*_{x_{max},x_{max}'}( \hat t_{max}) + \frac{\Phi^{-1}(\alpha) \hat \sigma}{c_N}\le \sup_{t \in C}\ell_{x_{max},x_{max}'}(t)\Big) \nonumber \\
    = & \mathbb{P}\Big(   \frac{c_N}{\hat \sigma}  \big( \ell_{x_{max},x_{max}'}( \hat t_{max})- \hat\ell_{x_{max},x_{max}'}^*(\hat t_{max})\big) \le \Phi^{-1}(\alpha) \Big) \nonumber \\
    & +o(1). \nonumber
\end{align}
In the second equality we have used the decomposition \eqref{Eq_B_decomposition}, together with the fact, that $B_1, B_2 = o_P(1)$. We can plug in the definition of the process $G$ into the probability on the right of \eqref{Eq_level_proof_1}, which gives us 
\begin{align}\label{Eq_Berry_1}
& \mathbb{P}\Big(   \frac{G(\hat t_{max})}{\hat \sigma}  \le \Phi^{-1}(\alpha) \Big) \\
=& \mathbb{P}\Big(   S(\hat t_{max})   \le \Phi^{-1}(\alpha) \Big) +o(1). \nonumber
\end{align}
Here we have used the definition of $S$ in \eqref{Eq_def_S}, as well as the (above mentioned) identity $
G(t)/\hat \sigma = S(t) + o_P(1),
$
which holds uniformly in $t \in U_\zeta(\mathcal{M})$ (recall that $\hat t_{max} \in \mathcal{M}$ with probability converging to $1$ according to \eqref{Eq_hat_t_in_U}).
Moreover, we have strictly speaking used that $S$ has (asymptotically)  a  continuous distribution function (see below).
Now recall that $\hat t_{max}$ (which is based on the samples $X_1,...,X_n$ and $Y_1,...,Y_n$ from the first step of the algorithm) is independent of all $X_1^*,...,X_N^*, Y_1,...,Y_N^*$ (and so loosely speaking of the randomness in $\tilde Z_i(\cdot)$). Thus we can express
\begin{align} \label{Eq_Berry_2}
     &\mathbb{P}\Big(   S(\hat t_{max})   \le \Phi^{-1}(\alpha) \Big) \\
    = & \int \mathbb{P}\Big(   S(t)   \le \Phi^{-1}(\alpha) \Big) d P^{\hat t_{max}}(t),\nonumber
\end{align}
where $P^{\hat t_{max}}$ is the image measure of $\hat t_{max}$.
Again we use that asymptotically the probability that $\hat t_{max} \not \in U_{\zeta}(\mathcal{M})$ converges to $0$ (see \eqref{Eq_hat_t_in_U}).  Now adding and substracting $\alpha$ yields 
\begin{align} \label{Eq_Berry_3}
& \alpha +o(1)\\
& +\int_{U_{\zeta}(\mathcal{M})} \mathbb{P}\Big(   S(t)   \le \Phi^{-1}(\alpha) \Big) - \alpha \,\, d P^{\hat t_{max}}(t) \nonumber\\
= & \alpha + o(1)\nonumber\\
&+\mathcal{O}\Big( \sup_{t \in U_{\zeta}(\mathcal{M})}\big|\mathbb{P}\Big(   S(t)   \le \Phi^{-1}(\alpha) \Big) - \Phi(\Phi^{-1}(\alpha)) \big|\Big). \nonumber
\end{align} 
Given some fixed $t$, the sum $S$ consists of i.i.d. random variables with unit variance and expectation $0$. We can therefore apply the Berry-Esseen theorem to see that 
$$
\sup_{t \in U_{\zeta}(\mathcal{M})}\big|\mathbb{P}\Big(   S(t)   \le \Phi^{-1}(\alpha) \Big) - \Phi(\Phi^{-1}(\alpha)) \big| = o(1),
$$
if we can show that (uniformly in $t$)
$$
\frac{\mathbb{E}|\tilde Z_1(t)-\mathbb{E}\tilde Z_1(t)|^3}{\sqrt{N}} =o(1).
$$
Similar calculations as before show that 
$$
\mathbb{E}|\tilde Z_1(t)-\mathbb{E}\tilde Z_1(t)|^3 = \mathcal{O}(h_{max}^{-1/2}),
$$
which proves the approximation and thus entails that \eqref{Eq_Berry_3} equals $\alpha +o(1)$. This again implies by \eqref{Eq_level_proof_1}, \eqref{Eq_Berry_1}, that the weak convergence in \eqref{Eq_level} holds and thus Theorem \ref{theorem_3} part i).\\

\end{document}